\documentclass[10pt]{iopart}

\usepackage{amsfonts}
\usepackage{amssymb}
\usepackage{xfrac}
\usepackage{graphicx}
\usepackage{booktabs,array}
\usepackage{comment}
\usepackage[skip=2pt,font=footnotesize]{caption}
\usepackage{titlesec}
\usepackage[flushleft]{threeparttable}
\usepackage{color}
\usepackage[normalem]{ulem}
\usepackage{tablefootnote}

\newcommand{\gene}{\textsc{GENE}}
\newcommand{\sa}{$\hat{s}-\alpha$}
\newcommand{\rlti}{$R/L_{Ti}$}
\newcommand{\rlte}{$R/L_{Te}$}
\newcommand{\rlne}{$R/L_{ne}$}
\newcommand{\exb}{$E{\times}B$}
\def\Apar{A_{\parallel}}
\def\vpar{v_{\parallel}}

\def\kpar{k_{\parallel}}

\newcolumntype{L}[1]{>{\raggedright\arraybackslash}m{#1}}
\newcolumntype{C}[1]{>{\centering\arraybackslash}m{#1}}
\newcolumntype{R}[1]{>{\raggedleft\arraybackslash}m{#1}}

\titleformat*{\section}{\normalsize\bfseries\rmfamily}
\titleformat*{\subsection}{\normalsize\bfseries\rmfamily}
\titleformat*{\subsubsection}{\normalsize\bfseries\rmfamily}

\graphicspath{{./Figures/}}
\DeclareGraphicsExtensions{.eps,.pdf,.jpg,.png}

\usepackage{verbatim}
\usepackage{caption}
\usepackage{subcaption}
\usepackage{dcolumn}
\usepackage{bm}

\usepackage[backend=bibtex,maxcitenames=10,sorting=none,style=ieee,natbib=true]{biblatex}
\bibliography{JCbibfile}

\begin{document}

\title{Integrated modelling and multiscale gyrokinetic validation study of ETG turbulence in a JET hybrid H-mode scenario}

\author[J Citrin$^{1,2}$, S Maeyama$^3$, C Angioni$^4$, N Bonanomi$^4$, C Bourdelle$^5$, F.J Casson$^6$, E Fable$^4$, T Goerler$^4$, P Mantica$^7$, A Mariani$^7$, M Sertoli$^{4,6}$, G Staebler$^8$, T Watanabe$^3$ and JET Contributors]{J Citrin$^{1,2}$, S Maeyama$^3$, C Angioni$^4$, N Bonanomi$^4$, C Bourdelle$^5$, F.J Casson$^6$, E Fable$^4$, T Goerler$^4$, P Mantica$^7$, A Mariani$^7$, M Sertoli$^{4,6}$, G Staebler$^8$, T Watanabe$^3$ and JET contributors$^*$}

\address{$^1$ DIFFER - Dutch Institute for Fundamental Energy Research, Eindhoven, the Netherlands}
\address{$^2$  Science and Technology of Nuclear Fusion Group, Eindhoven University of Technology, Eindhoven, Netherlands}
\address{$^3$ Department of Physics, Nagoya University, Nagoya 464-8602, Japan}
\address{$^4$ Max Planck Institute for Plasma Physics, Boltzmannstr. 2, 85748 Garching, Germany}
\address{$^5$ 
CEA, IRFM, F-13108 Saint Paul-lez-Durance, France}
\address{$^6$ CCFE, Culham Science Centre, Abingdon, OX14 3DB, United Kingdom of Great Britain and Northern Ireland}
\address{$^7$ Department of Physics “G. Occhialini”, University of Milano-Bicocca, Institute for Plasma Science and Technology, CNR, Milano, Italy}
\address{$^8$ General Atomics, P.O. Box 85608 San Diego, California 92121, USA}
\address{$^*$ See the author list of ‘Overview of JET results for optimising ITER operation’ by J. Mailloux et al. to be published in Nuclear Fusion Special issue: Overview and Summary Papers from the 28th Fusion Energy Conference (Nice, France, 10-15 May 2021)}

\date{\today}

\begin{abstract}
Previous studies with first-principle-based integrated modelling suggested that ETG turbulence may lead to an anti-GyroBohm isotope scaling in JET high-performance hybrid H-mode scenarios. A dedicated comparison study against higher-fidelity turbulence modelling invalidates this claim. Ion-scale turbulence with magnetic field perturbations included, can match the power balance fluxes within temperature gradient error margins. Multiscale gyrokinetic simulations from two distinct codes produce no significant ETG heat flux, demonstrating that simple rules-of-thumb are insufficient criteria for its onset.

\end{abstract}

\maketitle
\ioptwocol



\section{Introduction} \label{Introduction}
An accurate predictive model for tokamak turbulent transport is a vital component of integrated tokamak simulation~\cite{poli:2018}. It enables physical interpretation of present-day experiments, scenario optimization, extrapolation to future scenarios and devices, and experimental design. The gyrokinetic framework has proven to be successful in quantitatively describing tokamak core turbulence~\cite{garbet:2010,white:2019}. For sufficient tractability for application within integrated modelling suites, reduced-order-models applying the quasilinear approximation with saturation levels tuned to nonlinear gyrokinetic simulations have been developed, such as QuaLiKiz~\cite{bourdelle:2015,citrin:2017} and TGLF~\cite{staebler:2007}. While these models reproduce measured tokamak core thermodynamic radial profiles across wide regimes, continuous comparison against both experiments and higher fidelity simulation is necessary for validation of the models and their predictions for future scenarios and devices. This paper focuses on the validation of the QuaLiKiz model for predictions of Electron Temperature Gradient (ETG) driven turbulence in a specific JET scenario. 

ETG turbulence is driven by modes on the electron Larmor-radius scale-length, where the ion response is essentially adiabatic due to finite Larmor radius effects. The relevant regimes in which ETG turbulence significantly contributes to electron heat transport is an open question. Gyrokinetic simulations on electron Larmor-radius scales have long showed the possibility that in spite of the small intrinsic ETG mode scale, extended radial streamers can provide experimentally relevant ETG mode driven electron heat fluxes~\cite{dorland:2000,jenko:2002}. Experimentally, dedicated experiments have shown evidence of sharp electron temperature gradient thresholds consistent with ETG microturbulence~\cite{smith:2015,ryter:2019}, and power-balance electron heat flux which cannot be reconciled in nonlinear gyrokinetic simulations by ion Larmor-radius scale (henceforth referred to as "ion-scale") fluxes~\cite{bonanomi:2018,mariani:2019,mantica:2021}. In single-scale nonlinear ETG simulations with no ion-scale eddies, electron-scale zonal flows saturate ETG turbulence, leading to a collisionality scaling for the saturated ETG electron heat flux amplitude~\cite{colyer:2017}. In nonlinear multiscale simulations with a realistic ion to electron mass ratio, a complex picture has emerged, involving cross-scale interactions~\cite{maeyama:2015}. ETG modes can weaken ion-scale zonal flows, increasing ion and electron transport on ion-scales. Ion-scale eddies can also suppress electron-scale streamers, quenching ETG electron heat transport in spite of linear ETG instability. Additional studies have shown that the cross-scale ETG quench depends on the strength of the ion-scale (ITG in the studies carried out) instability drive; for sufficiently low ion-scale drive, ETG streamers survive and can produce significant electron heat flux, explaining experimental observations in CMOD and DIII-D~\cite{howard:2014c,howard:2016a,holland:2017}. These simulations have led to proposed rules-of-thumb where significant ETG fluxes in multiscale simulations can be expected. In Ref.~\cite{howard:2016b}, the ratio $\gamma_{high-k}/\gamma_{low-k}\ge40$ was correlated with significant cross-scale coupling. Refs.~\cite{staebler:2017,creely:2019} suggest the following as a necessary criteria for significant ETG fluxes:
\begin{equation}
\label{eq:thumbgary}
\frac{\gamma}{k_y}|_\mathrm{high-k} > \frac{\gamma}{k_y}|_\mathrm{low-k}
\end{equation}
where $\gamma$ is the mode growth rate, $k_y$ the binormal wavenumber, $\mathrm{high-k}$ corresponds to ETG scales ($0.1<k_y\rho_e<1$) and $\mathrm{low-k}$ ($0.1<k_y\rho_i<1$) corresponds to ion-scales, where $\rho_{e,i}$ is the electron/ion Larmor radius. For deuterium, $\rho_i/\rho_e\sim60$, being the square root of the ion to electron mass ratio. At each scale the peak value of the ratios are taken. $MAX\left(\frac{\gamma}{k_y}|_\mathrm{low-k}\right)$ is related to the zonal RMS {\exb} velocity under the hypothesis of a zonal flow $k_x$ mixing saturation mechanism~\cite{staebler:2016}. Additional hidden variables were not ruled out. The extreme computational expense of multiscale simulations, typically $10^7$ CPUh for a single converged simulation, precludes extensive parameter variations for testing the rules' robustness. 

A counter-example to a literal interpretation of the rule-of-thumb was recently shown in analysis of dedicated JET experiments~\cite{mantica:2021}. In that work, high stiffness in the electron heat channel was experimentally observed, with insufficient electron heat flux predicted in ion-scale simulations. However, multiscale simulations with realistic input parameters did not produce significant ETG electron heat flux in spite of the linear calculations passing the threshold in Eq.~\ref{eq:thumbgary}. 

Quasilinear turbulence models can incorporate such multiscale interactions into their saturation rules. QuaLiKiz has a simple rule, where ETG flux is quenched when the growth rate ratio between the peaks observed at electron -scales $0.1<k_y\rho_e<1$ and ion-scales $0.1<k_y\rho_i<1$, does not exceed the square root of the ion to electron mass ratio~\cite{citrin:2017}. TGLF employs a more refined zonal-flow mixing model to reproduce the phenomenology of multiscale nonlinear simulations~\cite{staebler:2016}. The multiscale saturation rule was employed in a multi-hierarchy TGLF validation study against CMOD and AUG discharges pointing to ETG electron heat flux as a necessary component in half the cases~\cite{creely:2019}. The importance of ETG flux is also predicted for electron-heated DIII-D ITER baseline discharges~\cite{grierson:2018}. 

This paper focuses on ETG predictions in the JET hybrid H-mode scenario (henceforth referred to as the `hybrid scenario') which is one of two main operational scenarios being developed for the DT campaign. The hybrid scenario operates in a regime with improved confinement compared to the $H_\mathrm{98}$ scaling law at relatively reduced plasma current and density, with a tailored q-profile and increased MHD stability allowing operation at higher $\beta_N$~\cite{joffrin:2005,hobirk:2012,challis:2015}. Numerous mechanisms contribute to the improved confinement, including increased $T_i/T_e$, increased NBI source penetration, an optimized q-profile (broad inner $q\sim1$ region with increased magnetic shear in the outer half-radius), increased {\exb} rotation shear, thermal and suprathermal-ion-enhanced electromagnetic (EM) stabilization of ITG, and increased Shafranov shift with virtuous core-edge coupling~\cite{garcia:2015,citrin:2015b}.

The improved ion heat confinement and typical $T_i/T_e>1$ in hybrid scenarios are expected to increase the potential importance of ETG turbulence; on the one hand through reduced ITG drive (see Eq.~\ref{eq:thumbgary}), and on the other hand through a decreased ETG critical gradient threshold, due to the $R/L_{Te}|_\mathrm{crit}\propto\left(1+Z_\mathrm{eff}\frac{T_e}{T_i}\right)$ dependence~\cite{jenko:2001b}. In a recent multi-channel integrated modelling study devoted to heavy impurity prediction and control in the JET hybrid scenario~\cite{casson:2020}, QuaLiKiz was applied for turbulent transport predictions within the JINTRAC~\cite{cenacchi:1988,romanelli:2014} integrated modelling suite. There, a significant role of ETG transport was predicted, with striking predictions in an isotope scan. The $T_e$ profile was pinned to near the ETG critical threshold, regardless of main ion isotope. Then, with increasing isotope mass, the decreasing ion-electron heat exchange enabled $T_i$ to increase, sustaining larger $T_i/T_e$ and hence an increased ITG instability threshold, improving confinement and increasing $T_i$ further, leading to an anti-GyroBohm ion mass confinement scaling. The larger $T_i/T_e$ also decreases the ETG critical threshold, leading to a reduction of $T_e$, further increasing $T_i/T_e$ and ITG stability. Without ETG, those feedback loops are not present and $T_i/T_e$ is predicted closer to 1, with higher $T_e$ and lower $T_i$. The mass scaling of the ion-electron heat exchange does not significantly impact the predicted confinement in the non-ETG case. We note that the simulations scanning isotope mass both with and without ETG maintained the same pedestal parameters, fast ion distributions, and rotation, which may all impact isotope mass confinement scaling in general. Furthermore, the predicted anti-GyroBohm isotope scaling owing to parallel electron dynamics~\cite{belli2020} is expected to play less of a role in ITG dominated H-modes, and in any case is not captured by QuaLiKiz. 

The potential benefit of ETG turbulence for DT scenario extrapolations strongly motivates deeper validation of these predictions. This is the focus of the present paper. Since the JET discharge in Ref.\cite{casson:2020} did not have core $T_i$ measurements, a more recent hybrid scenario was selected for analysis. The rest of the paper is as follows: section~\ref{sec:discharge} describes the JET discharge chosen for analysis and integrated modelling data preparation; section~\ref{sec:intmodelling} describes the JINTRAC-QuaLiKiz integrated modelling of this discharge, and sensitivity studies to physics assumptions and QuaLiKiz version; section~\ref{sec:GKmodelling} describes the linear and nonlinear single-scale and multiscale gyrokinetic modelling of the discharge at a single radial point of interest, serving as a validation of QuaLiKiz predictions and a test of its assumptions; conclusions are provided in section~\ref{sec:conclusions}. While this work focuses on ETG predictions and validation, including the ramifications of ETG for isotope confinement scaling, a related work has recently been published which studies - using integrated modelling -  the isotope scaling predictions of this discharge more generally~\cite{mariani:2021}.

\section{Characteristics of analyzed discharge and data preparation}
\label{sec:discharge}
We analyzed hybrid discharge \#94875 from the JET C38 deuterium campaign. Its basic parameters are listed in Table~\ref{tab:discharge}. $B_T$, $I_p$, and total input power are the same as in hybrid discharge \#92398 in Ref.~\cite{casson:2020}, whose modelled ETG predictions have motivated the validation exercise in this paper. $\beta_N$ is $\sim10\%$ lower in discharge \#94875. As opposed to \#92398, discharge \#94875 has high quality core $T_i$ measurements, provided by Neon Charge Exchange (Ne CX) obtained through (trace) Ne seeding.

\begin{table}
    \caption{Basic parameters of JET hybrid discharge \#94875 (pure deuterium), within the flattop time window analyzed. $B_T$ is the vacuum toroidal magnetic field at the magnetic axis. $I_p$ is the plasma current. $P_\mathrm{NBI}$ and $P_\mathrm{ICRH}$ are the Neutral Beam Injection (NBI) and Ion Cyclotron Resonance Heating (ICRH) total powers, respectively. $\beta_N\equiv\langle\beta\rangle\frac{aB_T}{I_p}$, where $a$ is the minor radius.}
    \centering
    \small
     \begin{tabular}{c|c|c|c|c|c|c|c}
	    $ B_T$ [T] & $ I_p$ [MA] & $ P_\mathrm{NBI} [MW] $ & $ P_\mathrm{ICRH} [MW] $ & $ \beta_N $  \\
        \hline
	    2.8 & 2.2 & 27 & 6.1 & 2.3 \\
	\end{tabular}
	\label{tab:discharge}
	\normalsize
\end{table}

\begin{figure*}[hbt]
	\centering
	\includegraphics[width=1.0\linewidth]{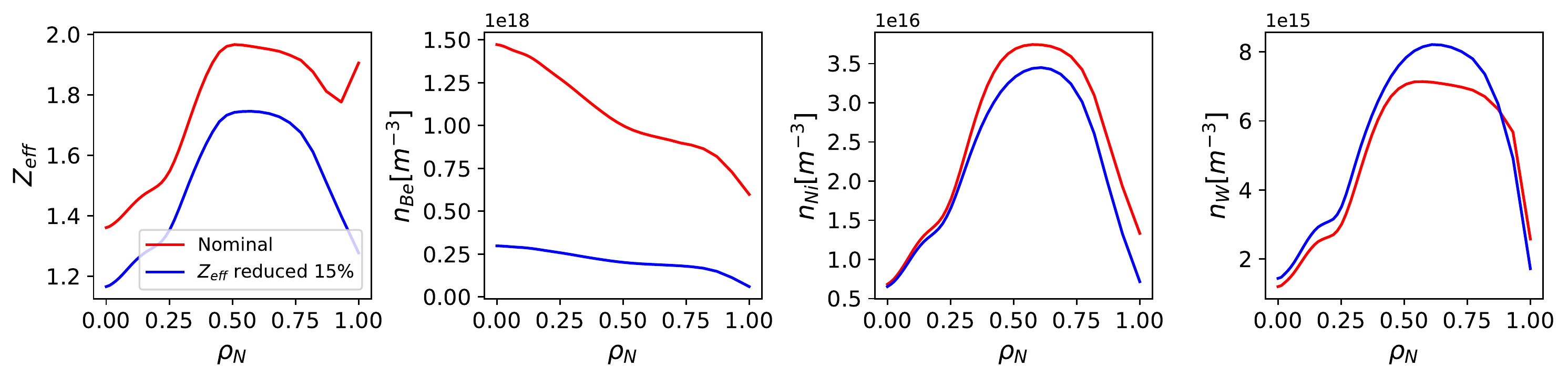}
	\caption{Flux-surface-averaged $Z_\mathrm{eff}$ and impurity densities as inferred from the method described in Ref.~\cite{sertoli:2019}. Both the nominal (red) and reduced within error bars (blue) $Z_\mathrm{eff}$ profiles are shown. The radiation inputs in the inference are the same in both cases.}
	\label{fig:sertoli_impurities}
\end{figure*}

A time window of $t=8.25-8.55~s$ was chosen for analysis, at the beginning of the stationary-state flattop following the density buildup after the L-H transition. All kinetic profiles, source and equilibrium data, were averaged within this window. $T_i$ and plasma rotation $v_\mathrm{tor}$ were measured with core and edge CX. $T_e$ and $n_e$ were measured with High Resolution Thomson Scattering (HRTS). Equilibrium reconstruction was carried out with EFIT++~\cite{lao:1985,appel:2018,szepesi:2020} with kinetic constraints. The initial q-profile applied within integrated modelling was calculated by EFIT++. The kinetic profile fits -- providing both initial condition and pedestal-top boundary conditions for the core integrated modelling -- were carried out with Gaussian Process Regression using EX2GK~\cite{ho:2019}, providing both fits and uncertainty envelopes. The NBI and ICRH (3.5\% H-minority) deposition profiles were calculated by a TRANSP~\cite{breslau:2018} interpretative simulation using NUBEAM~\cite{goldston:1981} and TORIC~\cite{brambilla:2002}. The impurity profiles for Be, Ni, and W, were inferred by a method incorporating multiple diagnostics for constrained consistency~\cite{sertoli:2019}. Due to the significance of $Z_\mathrm{eff}$ in setting ETG stability~\cite{jenko:2001b}, two sets of impurity profiles were generated, both nominal and propagating a 15\% reduction on line-averaged $Z_\mathrm{eff}$ (lower bound of error bar). The measured radiation characteristics (with total $P_\mathrm{rad}=9.2MW$) were kept constant through simultaneous modification of Ni and W. The two sets of inferred impurity profiles are shown in figure~\ref{fig:sertoli_impurities}. Sensitivity of ETG to $Z_\mathrm{eff}$ within integrated modelling is discussed in the next section. 

\begin{table*}
    \caption{Description of predicted and prescribed physical quantities in the integrated modelling simulations}
    \centering
    \small
      \begin{tabular}{c|c|c|c|c|c|c|c}
	    $n_{e,i}$ & $T_e$ & $T_i$ & $v_{tor}$ & $n_{imp}$ & $P_{rad}$ & $j$ & Equilibrium \\
	    \hline
	    \begin{tabular}{@{}c@{}}Predictive \\ (for $\rho<0.85$) \end{tabular} & \begin{tabular}{@{}c@{}}Predictive \\ (for $\rho<0.85$) \end{tabular} & \begin{tabular}{@{}c@{}}Predictive \\ (for $\rho<0.85$) \end{tabular} & Prescribed & \begin{tabular}{@{}c@{}}Prescribed \\ (reconstruction) \end{tabular} & \begin{tabular}{@{}c@{}}Prescribed \\ (bolometry) \end{tabular} & Predictive & \begin{tabular}{@{}c@{}}Prescribed \\ (EFIT++) \end{tabular}  \\

	\end{tabular}
	\label{tab:modelchoices}
	\normalsize
\end{table*}

\section{Results from integrated modelling}
\label{sec:intmodelling}

The first step in the validation exercise is to ascertain whether QuaLiKiz can reproduce the kinetic profiles of the deuterium discharge \#94875, and to investigate the predicted relevance of ETG turbulence. The QuaLiKiz quasilinear gyrokinetic turbulent transport model~\cite{bourdelle:2015,citrin:2017,stephens:2021} is electrostatic, and limited to {\sa} shifted circle geometry. Electromagnetic (EM) stabilization of ITG turbulence is taken into account through an ad-hoc model developed in Ref.~\cite{casson:2020}, whereby the QuaLiKiz {\rlti} input is locally reduced by the ratio of thermal to total (including suprathermal) pressure, $\frac{P_\mathrm{th}}{P_\mathrm{supra}+P_\mathrm{th}}$. This approximates the impact of both thermal and suprathermal contributions to $\beta$ stabilization of ITG.

Two versions of QuaLiKiz will be compared, QuaLiKiz 2.6.1 and QuaLiKiz 2.8.2. QuaLiKiz 2.6.1 is the version applied in Ref.~\cite{casson:2020}, applied here to determine whether the same trends are observed in modelling of discharge \#94875. These simulations are then compared to the most recent release 2.8.2 which includes two significant physics modifications compared to 2.6.1. Firstly, the Krook-like collision operator (ion-electron collisions for trapped electrons) was improved through comparison with higher-fidelity {\gene}~\cite{jenko:2000b} linear gyrokinetic simulations~\cite{stephens:2021b}. This improves the Trapped Electron Mode (TEM) dependence on collisionality, providing better density peaking predictions at mid to high values of collisionality, as well as increasing $Q_e$ predicted on ion-scales in general. Secondly, the ETG saturation rule was recalibrated by reducing the ETG saturation level by factor $1/3$ compared to Ref.~\cite{citrin:2017} following comparison with ETG multiscale {\gene} simulations from Ref.~\cite{bonanomi:2018}. In addition, the multiscale prefactor to the ETG saturation level was modified to be directly based on the $\left(\gamma/k\right)_\mathrm{max}$ at each spectral scale. The version 2.6.1 multiscale prefactor was:
\begin{equation}
    C_\mathrm{multiscale}=\frac{1}{1+e^{-\frac{1}{5}\left(\frac{\gamma_{\mathrm{maxETG}}}{\gamma_{\mathrm{maxITG}}}-\sqrt{\frac{m_i}{m_e}}\right)}}
\end{equation}
and in 2.8.2 has been modified to:
\begin{equation}
    C_\mathrm{multiscale}=\frac{1}{1+e^{-5\left(\frac{\left(\gamma/k\right)_{\mathrm{maxETG}}}{\left(\gamma/k\right)_{\mathrm{maxITG}}}-1\right)}}
\end{equation}
where the functional form is constructed such that this prefactor to the ETG flux approaches zero for $\left(\gamma/k\right)_{\mathrm{maxITG}}>\left(\gamma/k\right)_{\mathrm{maxETG}}$, and approaches unity for $\left(\gamma/k\right)_{\mathrm{maxETG}}>\left(\gamma/k\right)_{\mathrm{maxITG}}$, with a narrow transition zone between the two states. The linear eigenmodes on $\rho_e$ scales are unchanged. The modification only impacts the saturation level.

Table~\ref{tab:modelchoices} summarizes the list of predicted and prescribed physical quantities in the integrated modelling simulations discussed in the subsequent sections.

\subsection{Integrated modelling with QuaLiKiz 2.6.1}
QuaLiKiz 2.6.1 was run in JINTRAC for predictive simulations of discharge \#94875. The measured $T_e$, $T_i$, $n_e$, $v_\mathrm{tor}$, inferred $q$-profile, and inferred impurity content, were all set as initial conditions, corresponding to the averaged profiles in the $t=8.25-8.55$ time-window. $v_\mathrm{tor}$ and impurities were left fixed. Heat and particle transport (deuterium and electrons), as well as current diffusion, were predictive. The simulation was run for 1 plasma second (several confinement times), sufficient for the temperature and density profiles to reach stationary state. The magnetic equilibrium was kept fixed at the pressure constrained EFIT++ solution. The NBI+RF heat and NBI particle sources were prescribed from the TRANSP solutions. The radiation sink was prescribed from bolometry. The core boundary condition was taken at normalized toroidal flux coordinate $\rho_N=0.85$, just inside the pedestal top. The profiles were evolved only inside this boundary condition. Gas puff fuelling was neglected, since the penetration for $\rho_N<0.85$ is negligible. The role of gas puff is captured by the prescribed density pedestal top. Neoclassical transport was calculated by NCLASS~\cite{houlberg:1997}. For the inner core, $\rho_N<0.2$,  a patch for electron heat transport was prescribed, as $\chi_e = e^{(-\rho/0.15)^2}~[m/s^2]$. An exponential form was chosen to avoid any transport coefficient discontinuity, which may arise from a simple constant $\chi_e$ patch applied a limited radial range. The ad-hoc patch compensates for deep core anomalous transport not calculated by QuaLiKiz, needed to avoid spurious $T_e$ peaking in the on-axis region. The missing transport cannot be explained by neoclassical transport which is negligible for the electron heat channel, nor by sawteeth which are absent or infrequent in hybrid scenarios. Recent work has indicated that KBM modes with elongated mode structures may be responsible for transport in this localized region~\cite{kumar:2021}. These modes are challenging for QuaLiKiz to calculate due to both its electromagnetic and strongly ballooning eigenmode assumptions. In addition, non-local effects such as turbulence spreading could provide transport fluxes in this narrow region. Presently, QuaLiKiz does not incorporate any non-local effects.

\begin{figure*}[hbt]
	\centering
	\includegraphics[width=1.0\linewidth]{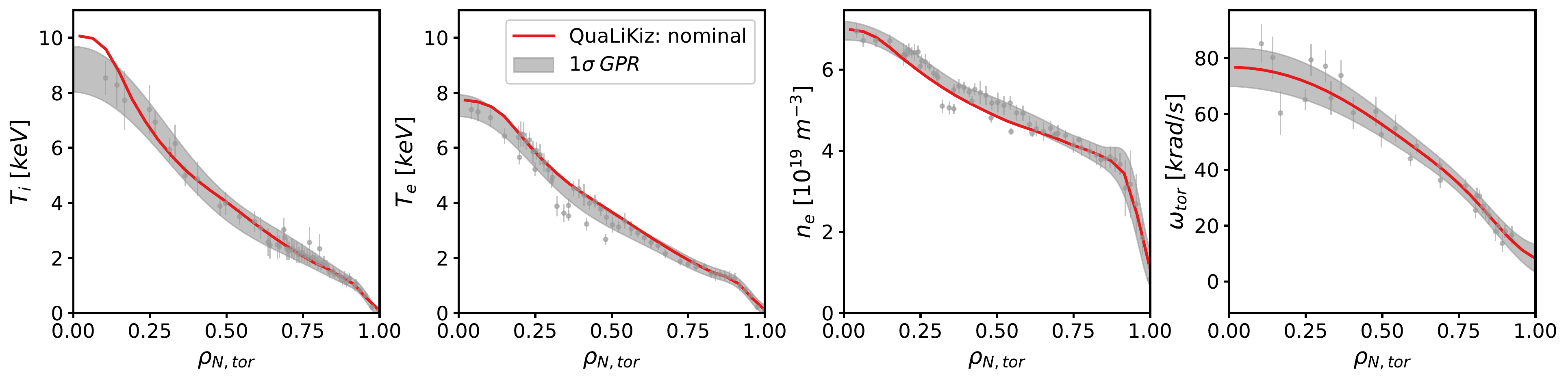}
	\caption{JINTRAC-QuaLiKiz\_2.6.1 nominal $Z_\mathrm{eff}$ simulation of JET hybrid scenario discharge \#94875. $T_i$, $T_e$, $n_e$ predictions within the $\rho_N=0.85$ boundary condition all agree with the Gaussian Process Regression (GPR) $1\sigma$ fit envelopes. The toroidal rotation (right panel) is prescribed}
	\label{fig:section3_nominal}
\end{figure*}

The integrated modelling results are shown in figure~\ref{fig:section3_nominal}, for the nominal $Z_\mathrm{eff}$ case. The predicted $T_e$, $T_i$ and $n_e$ profiles for $\rho_N<0.85$ are all within $1\sigma$ of the fitted profile uncertainty envelopes, constituting a successful reproduction of the scenario. 

\begin{figure}[hbt]
	\centering
	\includegraphics[width=1.0\linewidth]{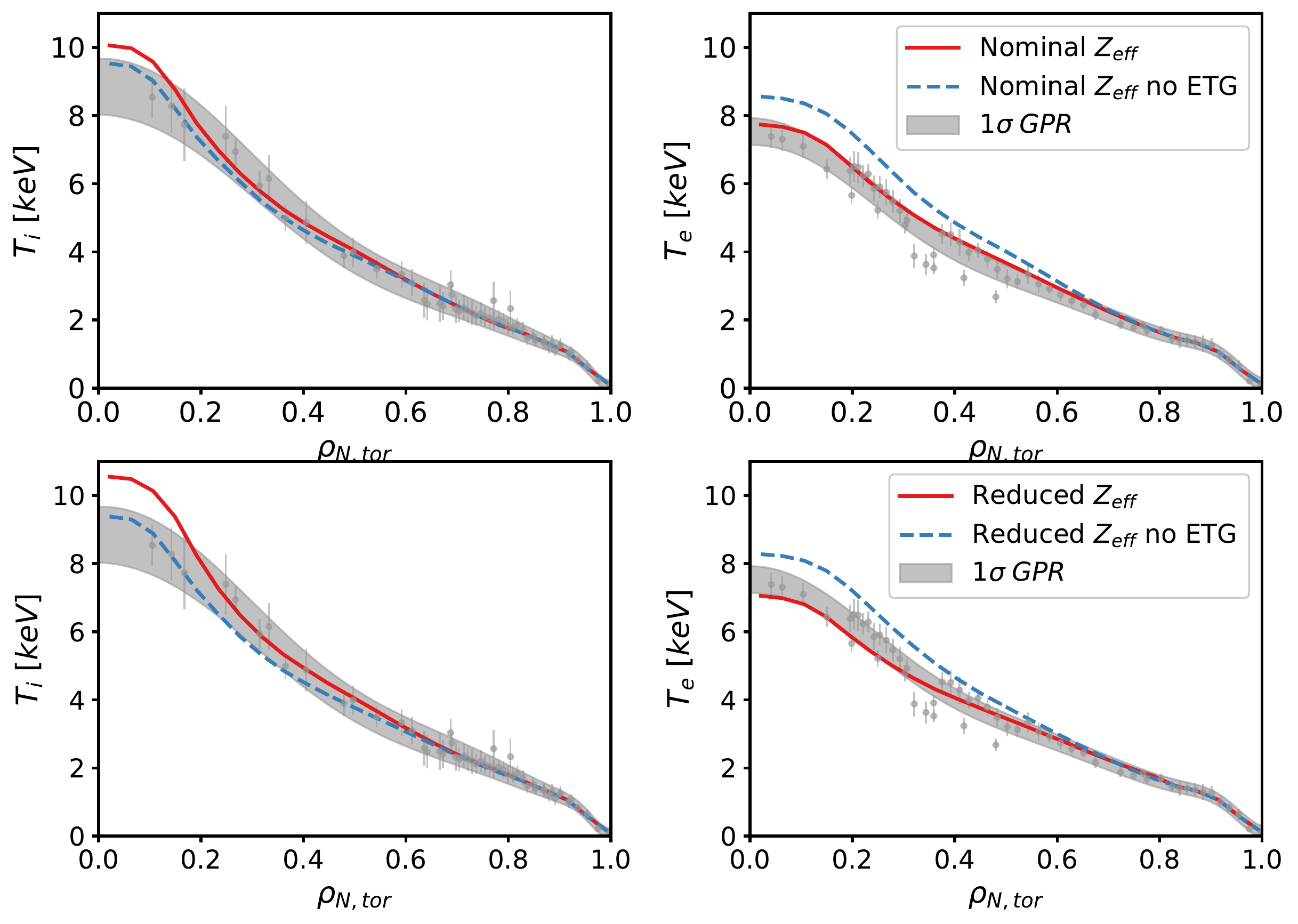}
	\caption{Comparison of JINTRAC-QuaLiKiz\_2.6.1 predictions with nominal (upper plots) and reduced (lower plots) $Z_\mathrm{eff}$, both with and without ETG. Only temperatures are shown for brevity}
	\label{fig:section3_Zeff}
\end{figure}

The impact of $Z_\mathrm{eff}$ and ETG are shown in figure~\ref{fig:section3_Zeff}. Since density predictions are similar for all cases, only $T_e$ and $T_i$ are shown for brevity. Without the inclusion of ETG turbulence, the predicted $T_e$ rises to levels above the GPR fit error envelope. Ion-scale turbulence alone - predicted by QuaLiKiz 2.6.1 to be ITG with a heat flux ratio of $Q_i/Qe\approx4$ in this case - cannot sustain the nominal base-case $Q_i/Q_e\approx2$ power balance ratio, which set by a combination of NBI power deposition (in this case ion dominated), ion-electron heat exchange (dependent on the $T_i - T_e$ temperature difference), and radiation. Therefore $T_e$ must rise to decrease the ion-electron heat coupling, and increase $Q_i/Q_e$. The clear impact of $Z_\mathrm{eff}$ on the importance of predicted ETG turbulence is observable through the increased $T_e$ difference between the with-ETG and no-ETG cases with 15\% reduced $Z_\mathrm{eff}$. Interestingly, $T_i$ is \textit{higher} for the cases with ETG, reminiscent of the anti-GyroBohm scaling of ion heat confinement in Ref.~\cite{casson:2020}, where ETG clamps $T_e$ and leads to increased $T_i/T_e$, improving ITG stability. The reduced $Z_\mathrm{eff}$ is within error bars. Its inclusion leads to predicted $T_e$ still within the 1$\sigma$ GPR uncertainty envelope. We therefore take the reduced $Z_\mathrm{eff}$ simulation with ETG as the ``base-case'' for subsequent sensitivity studies, and its output used for  validation against higher-fidelity gyrokinetic models as reported in section~\ref{sec:GKmodelling}.

\begin{figure}[hbt]
	\centering
	\includegraphics[width=0.8\linewidth]{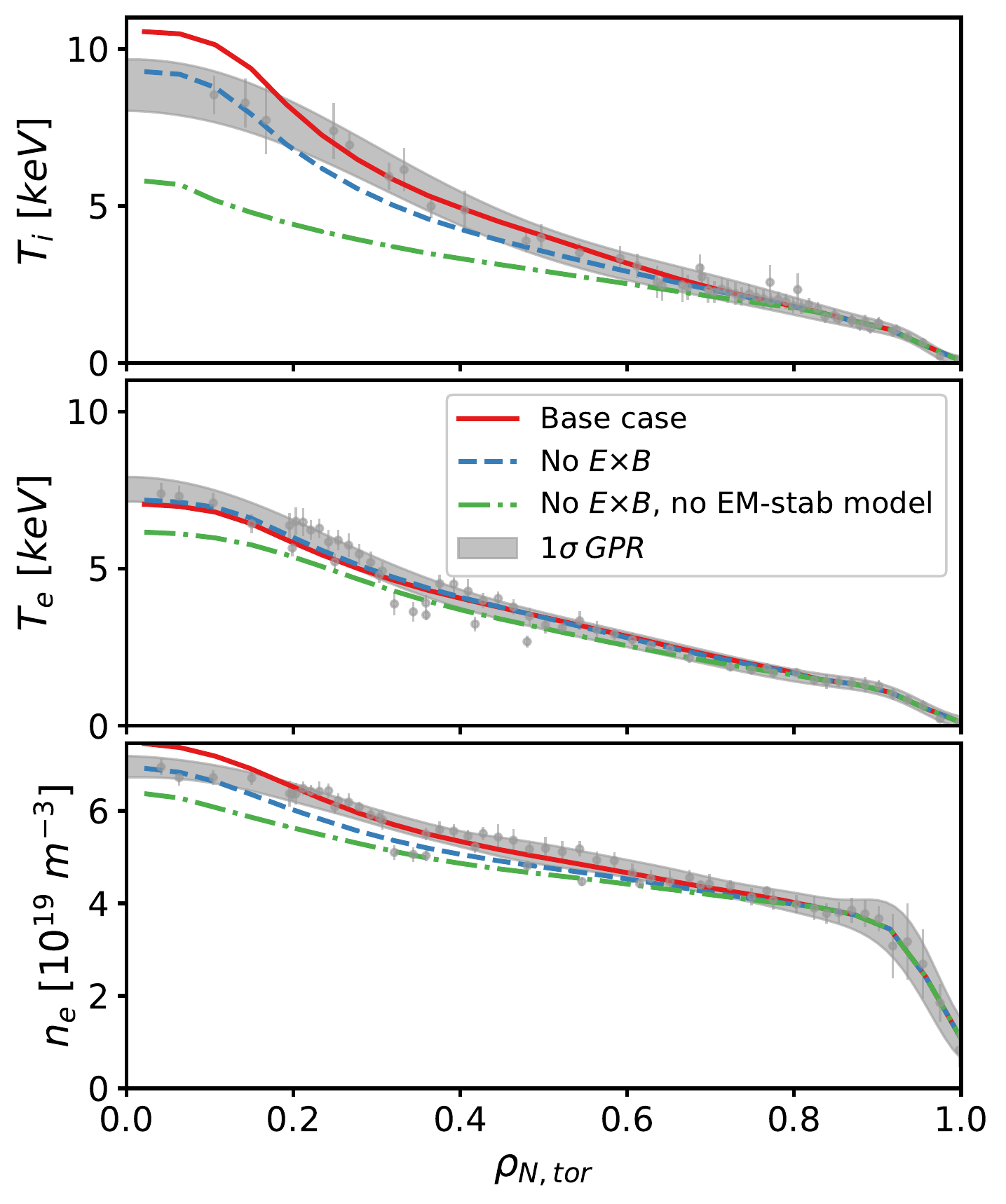}
	\caption{Comparison of JINTRAC-QuaLiKiz\_2.6.1 predictions between the base-case, a case with the {\exb} rotation shear turbulence suppression model turned off, and a case with both {\exb} suppression and the EM-stabilization models turned off, for $T_i$ (upper panel), $T_e$ (middle panel), and $n_e$ (lower panel) predictions}
	\label{fig:section3_sensitivity}
\end{figure}

The improved ion heat confinement regime typical of the hybrid scenario is predicted to rely on both {\exb} rotation shear and EM-stabilization, leading to higher $T_i/T_e$ and further improving ion heat confinement. This is illustrated in figure~\ref{fig:section3_sensitivity}. The base-case simulation is compared to a simulation where the QuaLiKiz {\exb} shear turbulence suppression model (see Refs.~\cite{cottier:2013,citrin:2017}) is turned off, and to a simulation where additionally the ad-hoc EM-stabilization model is turned off. {\exb} shear is important in this scenario in the outer half radius, owing to the QuaLiKiz assumption of ignoring {\exb} in the inner half radius due to the underprediction of destabilizing parallel velocity gradient modes~\cite{citrin:2017}. More dominant is the importance of the ad-hoc EM-stabilization rule, which increases on-axis $T_i$ by $\sim50\%$. While EM-stabilization has been shown to be critical for attaining high ion heat confinement in hybrid scenarios~\cite{citrin:2015b}, including for this same discharge~\cite{mariani:2021}, we stress that the ad-hoc nature of the EM-stabilization rule in QuaLiKiz is a caveat, to be addressed in future work.

For the case without {\exb} and EM-stabilization, no ETG turbulence is predicted. This is a consequence of increased $T_e/T_i$ and hence increased ETG critical thresholds, underlining the importance of improved ion confinement for attaining an ETG turbulence regime. 

\subsection{Integrated modelling with QuaLiKiz 2.8.2}
\begin{figure}[hbt]
	\centering
	\includegraphics[width=1.0\linewidth]{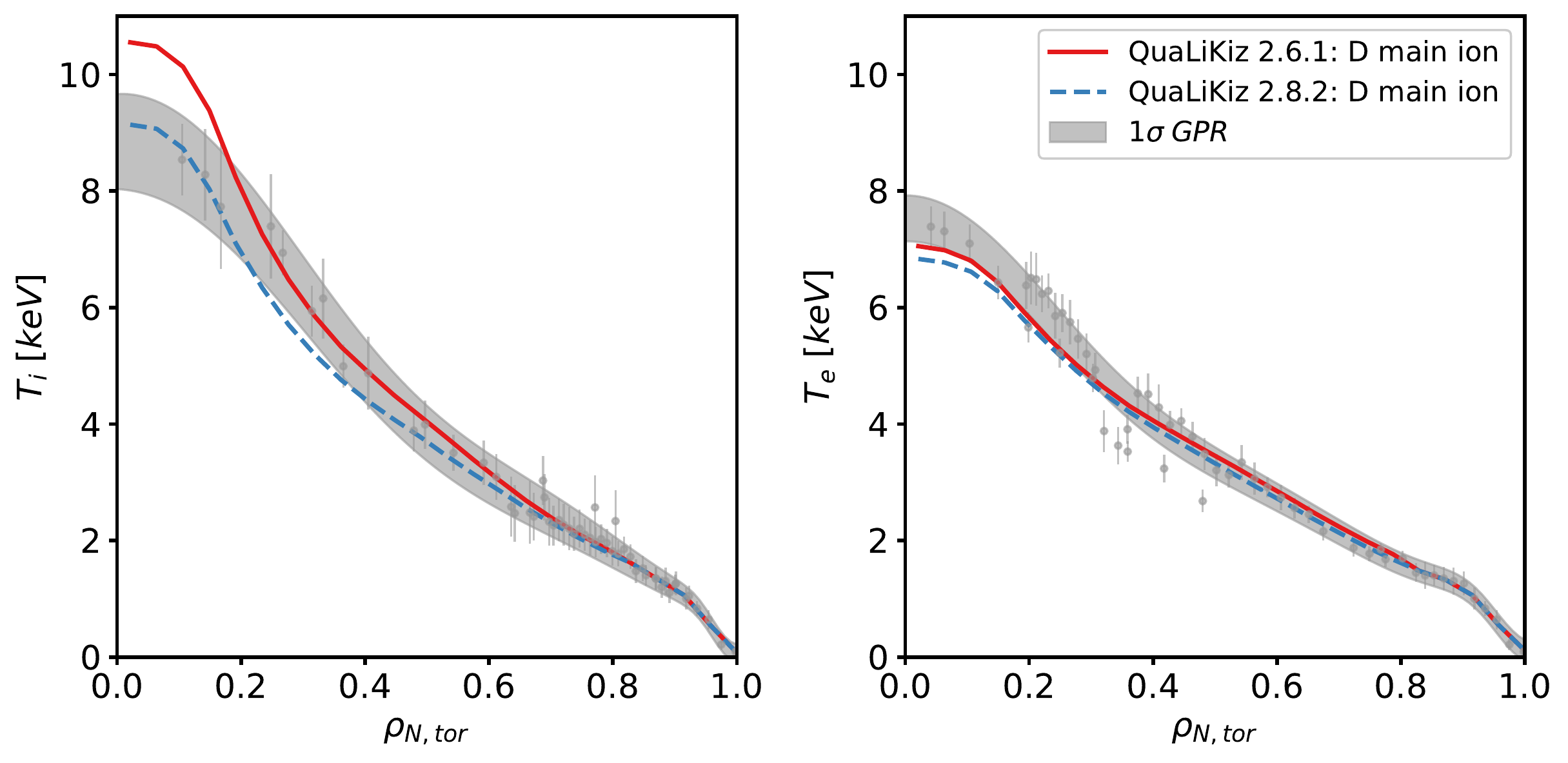}
	\caption{Comparison of JINTRAC-QuaLiKiz\_2.6.1 and JINTRAC-QuaLiKiz\_2.8.2 for the nominal Deuterium (D) main ion case. Predictions are shown for $T_i$ (left column) and $T_e$ (right column).}
	\label{fig:section3_versioncomp_extra}
\end{figure}

Modelling \#94875 with QuaLiKiz 2.6.1 indeed reproduces the same trends as in Ref.~\cite{casson:2020}. We now explore the differences obtained with the most recent QuaLiKiz release 2.8.2. Figure~\ref{fig:section3_versioncomp_extra} shows the comparison for the deuterium (D) main ion base-case between QuaLiKiz 2.6.1 and 2.8.2. Predictions for $T_e$ and $T_i$ are similar. While 2.8.2 predicts slightly lowered temperatures (particularly for $T_i$), they are still predominantly within the $1\sigma$ error envelope. $n_e$ is similar for both cases and omitted for brevity and clarity. 

\begin{figure}[hbt]
	\centering
	\includegraphics[width=1.0\linewidth]{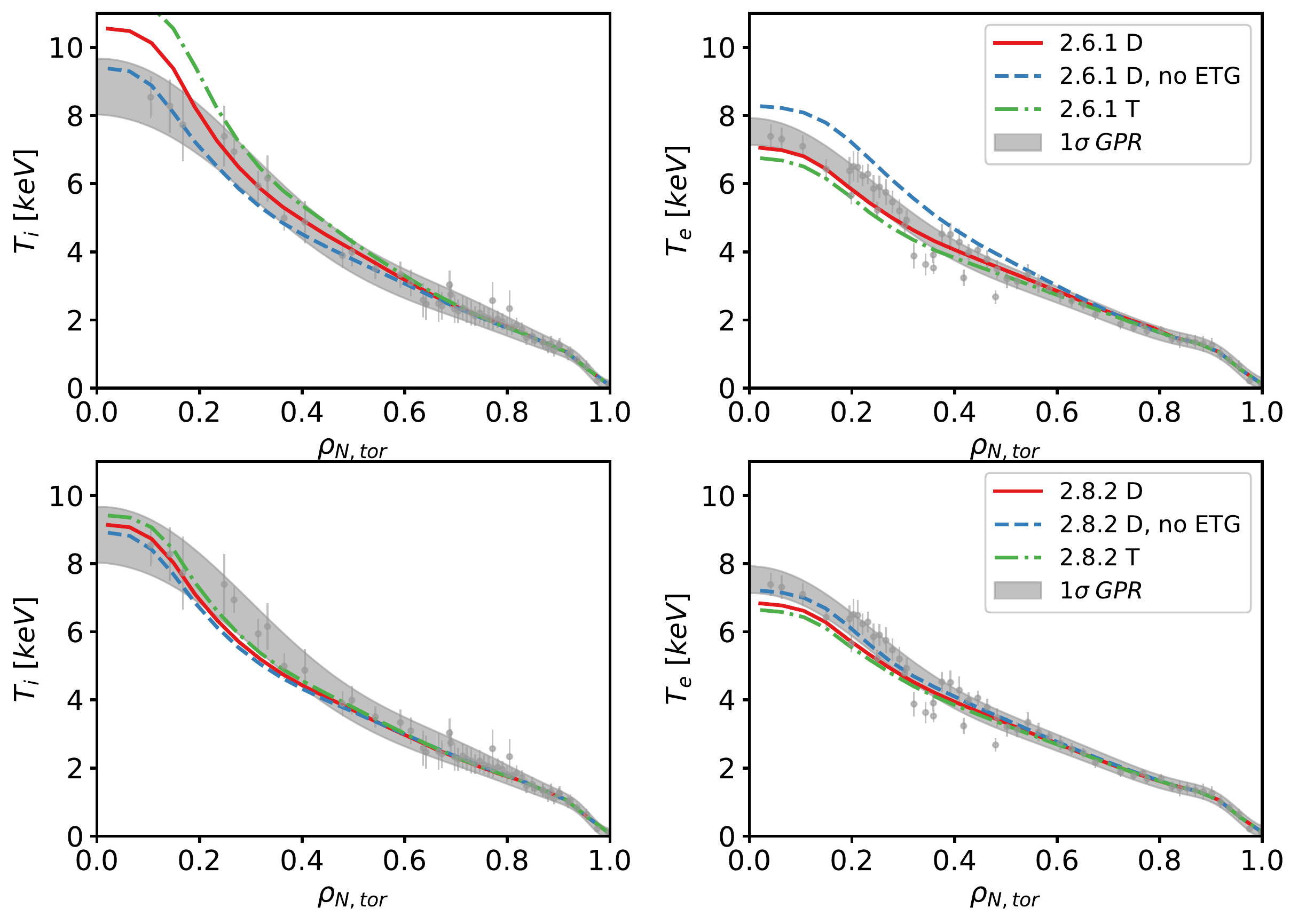}
	\caption{Comparison of JINTRAC-QuaLiKiz\_2.6.1 and JINTRAC-QuaLiKiz\_2.8.2 for ETG and isotope scaling predictions for discharge \#94875. The upper row is QuaLiKiz 2.6.1, the lower row 2.8.2. Predictions are shown for $T_i$ (left column) and $T_e$ (right column).}
	\label{fig:section3_versioncomp}
\end{figure}

\begin{figure}[hbt]
	\centering
	\includegraphics[width=1.0\linewidth]{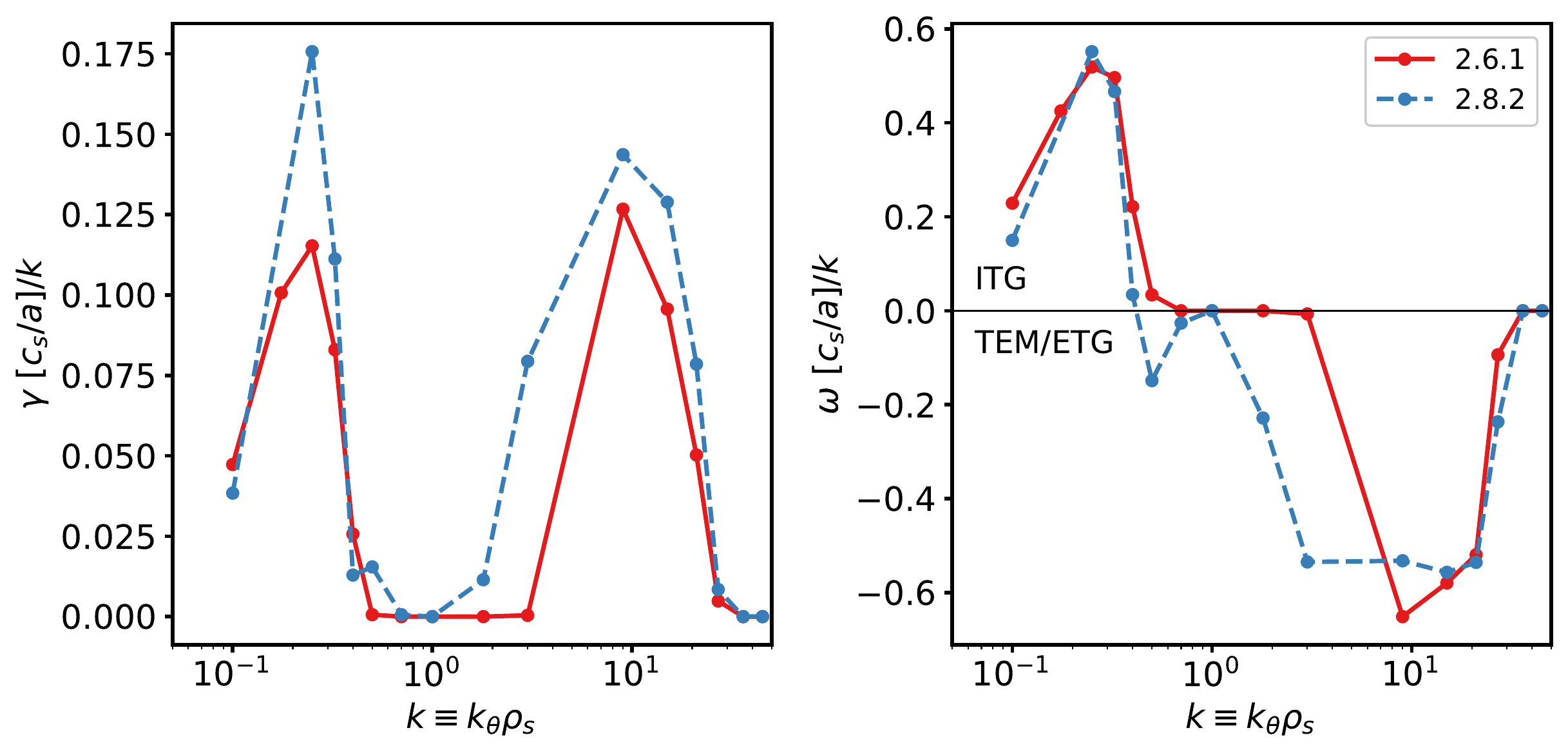}
	\caption{Comparison of JINTRAC-QuaLiKiz\_2.6.1 and JINTRAC-QuaLiKiz\_2.8.2 microstability predictions for the base-case integrated modelling run at $\rho=0.65$. The ratio of growth rate $\gamma$ to wavenumber $k$ is shown in the left panel. $\gamma$ is normalized by $c_s/a$, where $c_s\equiv\sqrt{T_e/m_i}$, and $a$ is the minor radius. $k$ is the poloidal wavenumber normalized by $1/\rho_s$, where the Larmor radius $\rho_s\equiv\sqrt{T_em_i}/qB$. The ratio of mode frequency $\omega$ (same normalization as $\gamma$) to wavenumber $k$ is shown in the right panel. The x-axis is logarithmic, to conveniently visualize both ion and electron scales.}
	\label{fig:section3_modecomp}
\end{figure}

While predictions both versions agree with the measurements for the base-case, the new QuaLiKiz version has a significant modification of the impact of predicted electron-scale turbulence and hence on the isotope scaling. As shown in Figure ~\ref{fig:section3_versioncomp}, three separate simulations were carried out for each QuaLiKiz version: with D as main ion and ETG scales included (the base-case as in Figure~\ref{fig:section3_versioncomp_extra}), with D as main ion and ETG scales removed, and with tritium (T) as main ion and ETG scales included. A striking observation is that removing ETG turbulence from the QuaLiKiz 2.8.2 predictions has a reduced impact compared to the 2.6.1 predictions. This translates to a significantly reduced anti-GyroBohm isotope scaling in the T simulations. For 2.8.2, all three simulations in the physics scan provided very similar results. An explanation is that the increased trapped electron drive in 2.8.2 due to the improved collision operator has increased the electron heat flux component of the ion-scale modes, reducing the necessity of electron-scale modes to supplement the electron heat flux. The fact that the improved collision operator in 2.8.2 is responsible for reduced impact of ETG in the integrated modelling, as opposed to the modified multiscale saturation rule in 2.8.2, was verified by running (not shown for brevity) a simulation using 2.8.2 with the 2.6.1 multiscale saturation rule, with similar results (both with and without ETG) to 2.8.2 itself. The impact of the improved collision operator on the linear modes is seen in figure~\ref{fig:section3_modecomp}, which shows a comparison of the QuaLiKiz linear growth rate and frequency predictions averaged over the last 100~ms of each respective base-case integrated modelling run, at $\rho=0.65$. For version 2.6.1, the ion-scale modes ($k_\theta \rho_s<2$) are purely ITG (positive frequencies - defined here as the ion diamagnetic direction). For version 2.8.2, TEM modes (negative frequencies in the electron diamagnetic direction) dominate the ion-scale spectra for $k_\theta \rho_s\geq0.5$. These contribute to the electron heat flux. Also the ITG modes in 2.8.2 display an increased $Q_e/Q_i$ compared to 2.6.1, further contributing to the ion-scale electron heat flux. In addition, in the 2.8.2 base-case simulation $T_i/T_e$ is lower, increasing the power balance $Q_i/Q_e$ ratio compared to the 2.6.1 case, easing the ability of ion-scale modes to provide the required flux, particularly when TEM plays a role. Figure~\ref{fig:section3_modecomp} clearly shows that $\left(\gamma/k\right)_\mathrm{ETG}<\left(\gamma/k\right)_\mathrm{ITG}$ at $\rho=0.65$ in the 2.8.2 simulation, while $\left(\gamma/k\right)_\mathrm{ETG}>\left(\gamma/k\right)_\mathrm{ITG}$ in 2.6.1, reflecting the relative importance of the role of ETG at the more outer radii in the simulations of the two different QuaLiKiz versions. At more inner radii ($\rho<0.5$), ETG is unstable and relevant in both base-case simulations regardless of version. 

To summarize, JINTRAC-QuaLiKiz simulations of JET hybrid discharge \#94875 using QuaLiKiz 2.6.1 agree with experimental measurements, and reproduce the phenomenology reported in Ref~\cite{casson:2020} regarding the importance of ETG turbulence and a mechanism for anti-GyroBohm isotope scaling. However, comparisons with the newer QuaLiKiz 2.8.2, which has increased trapped electron drive owing to an improvement of the collision operator, leads to a decreased importance of ETG while still maintaining agreement with the experimental measurements within $1\sigma$ for the bulk of the profiles. With 2.8.2, the anti-GyroBohm isotope scaling predictions are significantly diminished. In the next section we proceed to a validation of QuaLiKiz predictions against higher-fidelity gyrokinetic modelling.

\section{Gyrokinetic analysis}
\label{sec:GKmodelling}
In this section we validate QuaLiKiz 2.6.1 and 2.8.2 predictions against the higher-fidelity GENE gyrokinetic code, in its local, gradient-driven, $\delta{f}$ mode. Both linear (with initial value solver) and nonlinear {\gene} simulations were carried out. {\gene} is a Eulerian gyrokinetic code, evolving the perturbed particle distribution functions self-consistently with the Maxwell field equations. {\gene} works in field aligned coordinates, where $x$ is the radial coordinate, $z$ the parallel coordinate along the field line, and $y$ the binormal coordinate. All shown simulations are spectral in both the $x$ and $y$ directions. Both {\sa} (as in QuaLiKiz) and parameterized shaped Miller geometry~\cite{miller:1998} were employed, as detailed later in this section. Collisions in {\gene} were modelled using a linearised Landau-Boltzmann operator. Typical grid parameters were as follows: 24 point discretisation in the parallel direction, 48 points in the parallel velocity direction, and 15 magnetic moments, where parallel velocity box ranged between $[-3v_{Tj},+3v_{Tj}]$, with thermal velocity $v_{Tj}=\sqrt{2T_j/m_j}$ where $T_j$ is the background Maxwellian temperature, $m$ the species mass, and $j$ the species identifier. The upper end of the magnetic moment box was set at $9T_j/B_\mathrm{ref}$, with $B_\mathrm{ref}$ the reference magnetic field strength (on-axis). For the linear runs, $n_{kx}=31$ radial wavenumbers were included. For the ion-scale nonlinear runs, the perpendicular box sizes were $[L_x,L_y]\approx[100,125]$ in units of ion Larmor radii, with $[n_{kx}, n_{ky}]=[128,32]$ perpendicular wavenumbers included. The perpendicular grid parameters in the multiscale simulations are discussed in section~\ref{sec:multiscale}. Numerical convergence was verified for all grid dimensions, both in the linear and nonlinear simulations, through a detailed study of the impact of modifying grid resolutions. All simulations carried out were electrostatic, unless explicitly specified in sensitivity tests. 

All analysis was carried out for parameters corresponding to the final timeslice of the base-case integrated modelling simulation, at normalized toroidal flux coordinate $\rho=0.65$. This is a location where ETG was predicted to significantly contribute to the electron heat flux according to the JINTRAC-QuaLiKiz\_2.6.1 modelling. Moreover, this outward location is associated with lower-$\beta$, presumed to justify excluding electromagnetic (EM) ITG stabilization effects in the {\gene} modelling. The inclusion of EM effects in the nonlinear multiscale modelling would not be feasible due to the computational expense, particularly considering the long time-scale dynamics encountered when enhanced zonal flow coupling occurs~\cite{disiena:2021}. At inner radii for this same discharge, EM effects are found to play a key role~\cite{mariani:2021}. Modifications from the exact base-case parameters due to the desire to maintain power-balance relevant fluxes under the modelling assumptions (e.g. no rotation shear and EM effects) are mentioned where appropriate in the subsequent sections. See table~\ref{tab:singlescaleparams} for a list of the main dimensionless input parameters, comparing the nominal measured values with the base-case JINTRAC-QuaLiKiz\_2.6.1 predicted values used for the code comparison.

\begin{table*}
    \caption{Dimensionless input parameters for the {\gene} vs QuaLiKiz single-scale study, including comparison of the nominal measured values to the values from the final timeslice of the base-case integrated modelling simulation used for the code comparison}
    \centering
      \begin{tabular}{c|c|c|c|c|c|c|c|c}
	    Case & {\rlti} & {\rlte} & {\rlne} & $\hat{s}$ & $q$ & $T_i/T_e$ & $\alpha$ & $Z_\mathrm{eff}$ \\
        \hline
	    JINTRAC-QuaLiKiz\_2.6.1 (base-case) & 9.25 & 8.62 & 2.93 & 1.5 & 2.02 & 1.08 & 0.62 & 1.65 \\
	    Measured & 8.03  & 8.72 & 3.46 & 1.4 & 2.1 & 1.12 & 0.71 & 1.55 \\	    
	\end{tabular}
	\label{tab:singlescaleparams}
	\normalsize
\end{table*}

A further simplification of the modelling inputs was to bundle all impurity species into a single effective impurity. This single impurity was chosen with a charge, density, and density gradient, such that the main ion density and density gradient ($n_D/n_e$ and $R/L_{nD}$) is unaltered compared to the 3-impurity case. Comparison of single-effective-impurity and 3-impurity {\gene} simulations (not shown for brevity) constituted a difference of under 5\%, while saving significant computational time in the {\gene} simulations. The impact of fast ions on EM-stabilization of ITG is explored in dedicated linear simulations discussed in section~\ref{sec:salpha}.

\subsection{Single scale simulations with simplified parameters}
\label{sec:salpha}

QuaLiKiz (both versions) and {\gene} (linear and nonlinear) simulations are compared in gradient-driven simulations: nonlinearly on ion scales only and linearly in both scales separately. The GENE simulations are carried out in {\sa} geometry for a comparison with QuaLiKiz under similar geometry assumptions. Rotation is not included in the shown simulations, due to the induced non-stationary Floquet modes in {\gene} which hinders comparison. 

\begin{figure}[hbt]
	\centering
	\includegraphics[width=1.0\linewidth]{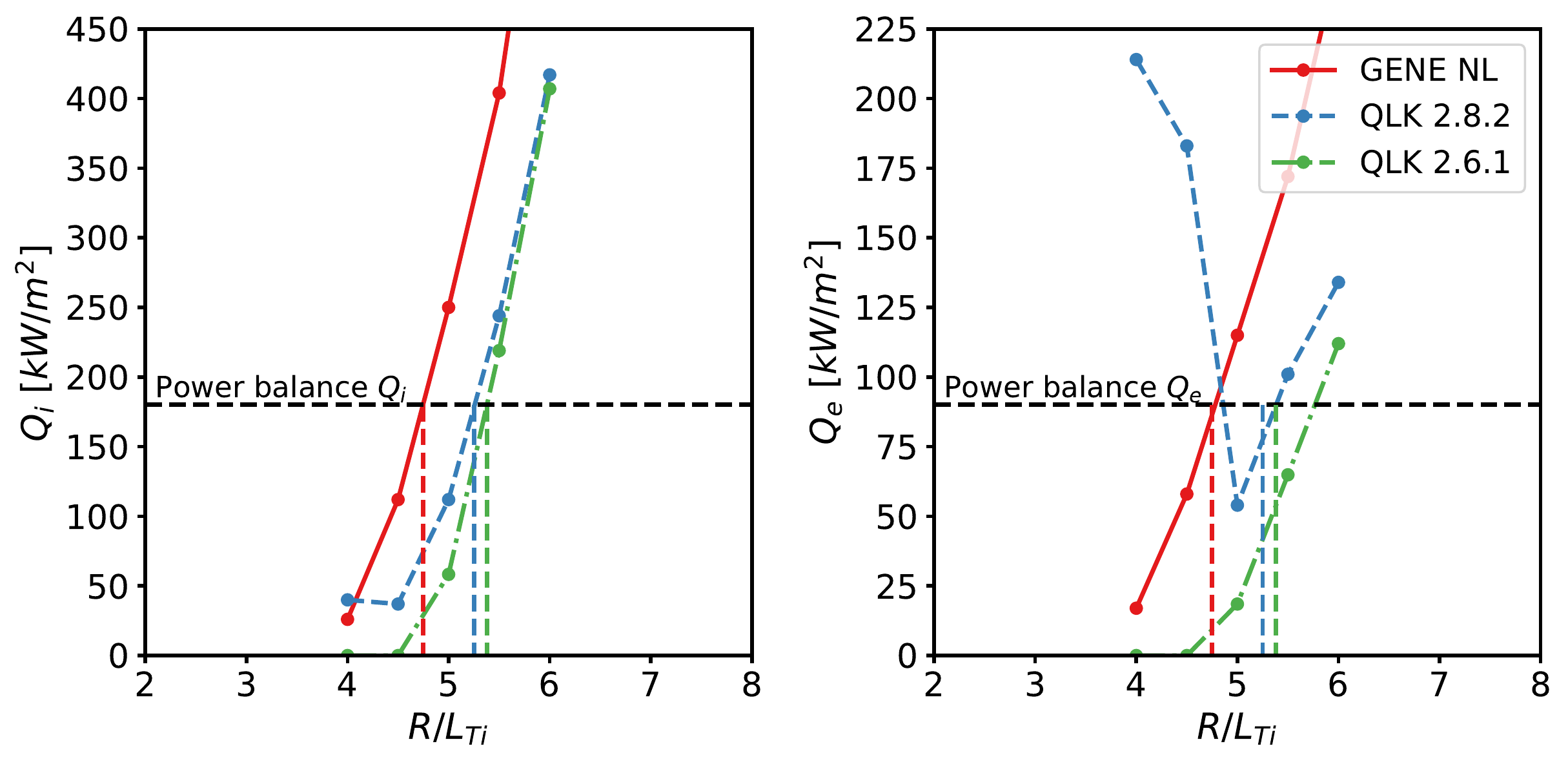}
	\caption{Comparison between JINTRAC-QuaLiKiz\_2.6.1 (green dashed curves), JINTRAC-QuaLiKiz\_2.8.2 (blue dashed curves), and nonlinear {\gene} (red solid curves) ion and electron heat flux predictions for the base-case integrated modelling run. The power balance heat fluxes from the integrated modelling run are portrayed by the horizontal black dashed lines. The red, blue, and green vertical dashed lines portray the $R/L_{Ti}$ corresponding to the $Q_i$ power balance levels for the {\gene}, QuaLiKiz 2.8.1, and QuaLiKiz 2.6.1 simulations respectively.}
	\label{fig:section4_QLKvsGENENL}
\end{figure}

An {\rlti} scan comparing nonlinear {\gene} and quasilinear QuaLiKiz fluxes is shown in figure~\ref{fig:section4_QLKvsGENENL}. The lack of rotation and EM-effects led to a power balance at an {\rlti} value significantly below the measured one ({\rlti}$\sim8$). A number of observations are evident from the figure:
\begin{itemize}
    \item {\gene} and QuaLiKiz agree on the base ITG stiffness level (sensitivity of ion heat flux to {\rlti})
    \item QuaLiKiz ITG thresholds are upshifted compared to {\gene} by {\rlti}$\approx0.5$ for this case, a relatively minor difference of 10\% in {\rlti}
    \item At the {\rlti} values corresponding to the ion heat flux power balance values (designated by the vertical lines), QuaLiKiz-2.8.2 and {\gene} agree on a heat flux ratio of $Q_i/Q_e\approx2$. On the other hand, QuaLiKiz-2.6.1 has a significantly lower electron heat flux on ion-scales with $Q_i/Q_e\approx4$. This underprediction of QuaLiKiz-2.6.1 ion-scale $Q_e$ is consistent with other analysis of this same discharge at lower radii: see Fig.9b in Ref.~\cite{mariani:2021}.
    \item At {\rlti} values below the power balance levels, QuaLiKiz-2.8.2 predicts a TEM dominated regime with $Q_e\gg Q_i$. This non-monotonic $Q_e$ is in disagreement with {\gene}. This arises from an over-prediction of TEM drive in QuaLiKiz in the low {\rlti} regime at these parameters. We stress that for this case, the disagreement occurs in a $Q_i$ range significantly below power-balance levels and thus does not impact the scenario prediction.
\end{itemize}

We can conclude - in the regime of physical interest at power balance levels - that QuaLiKiz-2.8.2 is better validated by the {\gene} comparison than QuaLiKiz-2.6.1. The increased electron heat flux is consistent with the increase in TEM drive due to the collision operator improvement, underpinning the reduced reliance in 2.8.2 on ETG to supplement the electron heat flux in the integrated modelling simulations. 

Regarding the impact of rotation, not shown for brevity, additional nonlinear-{\gene} and QuaLiKiz {\rlti} scans where carried out with {\exb} shear included. This had a similar impact for both models, leading to an {\rlti} upshift of $\approx1$. 

\begin{figure}[hbt]
	\centering
	\includegraphics[width=1.0\linewidth]{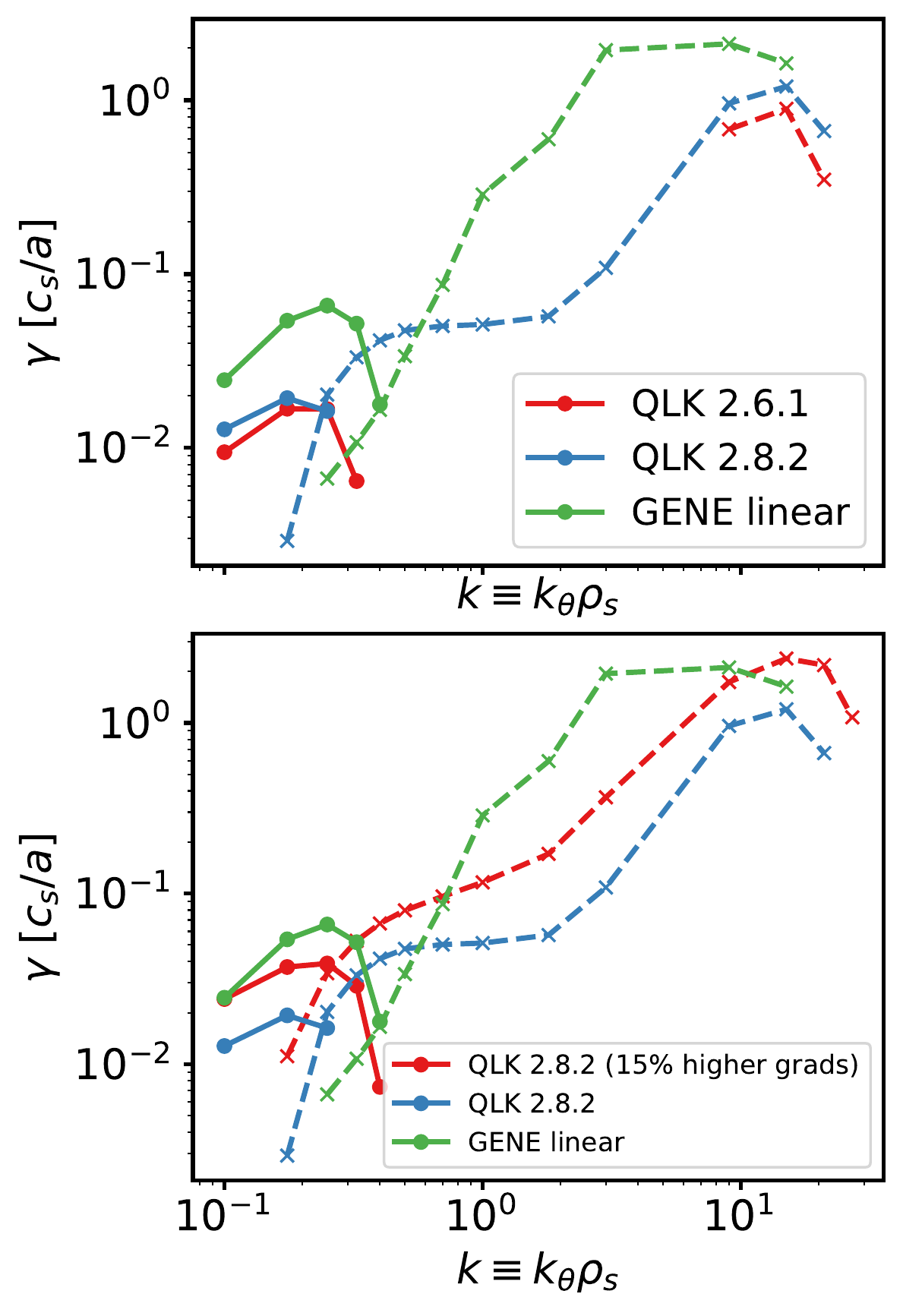}
	\caption{Comparison between QuaLiKiz and {\gene} linear spectra for the same parameters as in figure~\ref{fig:section4_QLKvsGENENL}, at {\rlti}=5.4. The upper panel compares QuaLiKiz-2.6.1 (red) QuaLiKiz-2.8.2 (blue), and linear {\gene} (green). Ion modes (ITG) are represented by the solid curves, and electron modes (TEM+ETG) by dashed curves. In the lower panel, QuaLiKiz-2.6.1 is absent and the red curves correspond to QuaLiKiz-2.8.2 for both {\rlti} and {\rlte} increased by 15\%}
	\label{fig:section4_QLKvsGENElin}
\end{figure}

The trends observed in figure~\ref{fig:section4_QLKvsGENENL} can be understood from examination of the underlying linear modes. This is shown in figure~\ref{fig:section4_QLKvsGENElin}. In the left panel, both QuaLiKiz-2.6.1 and 2.8.2 have reduced ITG growth rates compared to {\gene}. This growth-rate discrepancy is mostly resolved by a 15\% increase in {\rlti} (see right panel), consistent with the power-balance {\rlti} shift in figure~\ref{fig:section4_QLKvsGENENL}. QuaLiKiz-2.6.1 has no TEM (interpreted as electron modes on ion-scales), whereas QuaLiKiz-2.8.2 has TEM growth rates above {\gene}, consistent with the TEM-dominated regime at low {\rlti}. Finally, {\gene} ETG growth rates are significant, with maximum $\gamma/k_\mathrm{ETG}$ approximately twice maximum $\gamma/k_\mathrm{ITG}$. This suggests that multiscale simulations should lead to significant ETG flux. This is not necessarily borne out, as will be discussed in section~\ref{sec:multiscale}. The QuaLiKiz ETG growth rates are in line with GENE growth rates following a 15\% {\rlte} increase. The fact that seemingly significant growth rate discrepancies are resolved by relatively minor modifications in driving gradient is typical of the challenges faced by gradient-driven comparisons. 

This section concludes with a study of the impact of EM-stabilization, to judge \textit{a posteriori} whether EM effects can indeed be neglected at the relatively outward radius of $\rho=0.65$ chosen for analysis of this case. Four cases were studied with linear {\gene}, with parameters as in figure~\ref{fig:section4_QLKvsGENElin} but increased ITG drive ({\rlti}$=8.3$), close to the experimentally measured value, such that the EM-stabilization impact on the nominal parameters is clearer. The electrostatic (ES) case ($\beta=0$) is compared with the EM case with nominal $\beta=0.68\%$, and with the addition of fast ions from the NBI heating, assuming Maxwellian fast ions with parameters taken from the TRANSP-NUBEAM modelling (see table~\ref{tab:NBI}). $\alpha_\mathrm{MHD}$ is kept fixed at the nominal value throughout. The inclusion of EM effects provides a very significant stabilization, as evident from the drop in growth rates from the red to blue curves. At the lowest $k$, a mode of a microtearing type is then dominant, although this is not expected to lead to significant flux in nonlinear modelling in conjunction with ITG~\cite{doerk:2015}. The growth rates are further reduced with the inclusion of fast ions across much of the spectrum. However, for nominal fast ions, a fast-ion driven mode dominates the spectrum at the lowest $k$. Nonlinearly, such modes in JET hybrid scenarios are predicted to drive significant thermal and fast-ion transport, and thus in a self-organized state are expected to remain marginally subcritical through self-consistent profile modifications~\cite{citrin:2015b}, although flux-driven high-fidelity modelling is necessary to shed light on the precise system dynamics. A modest 20\% reduction in fast-ion parameters (density and gradients) suppresses this mode, as seen in the purple curve. 

This phenomenology is strongly reminiscent of the typical observations at more inner radii in hybrid scenarios, see e.g. Ref.~\cite{citrin:2015b}. While striking that it is observed here at outer radii, the observation is also strongly geometry dependent. As later analyzed in section~\ref{sec:miller}, the linear EM-stabilization effect is weakened with shaped geometry.  

\begin{table}
    \caption{Fast ion (D-NBI) species parameters for the linear {\gene} EM simulations shown in figure~\ref{fig:section4_EM}. Parameters were calculated from the TRANSP-NUBEAM interpretative simulation results prescribed in the predictive integrated modelling}
    \centering
    \small
     \begin{tabular}{c|c|c|c|}
	    $ n_{fast}/n_e$ & $T_{fast}/T_e$ [MA] & $ R/L_{Tfast}$ & $ R/L_{nfast}$ \\
        \hline
	    0.0434 & 11.24 & 3.81 & 11.24 \\
	\end{tabular}
	\label{tab:NBI}
	\normalsize
\end{table}

\begin{figure}[hbt]
	\centering
	\includegraphics[width=1.0\linewidth]{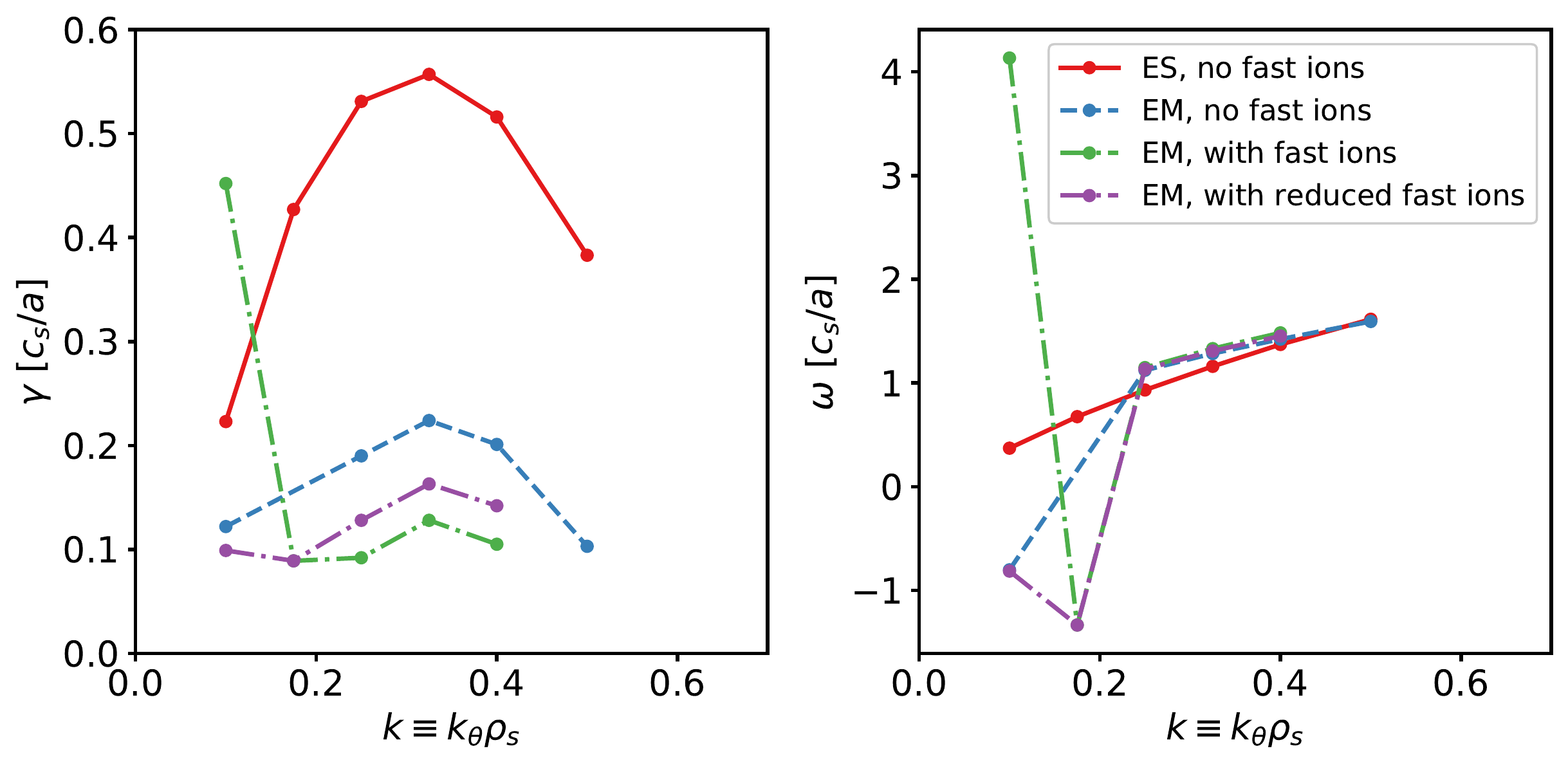}
	\caption{Study of the impact of EM-stabilization in {\gene} {\sa} linear runs. Parameters are the same as in figure~\ref{fig:section4_QLKvsGENElin}, but with {\rlti}={\rlte}=8.3. Ion-scale growth rate spectra are compared for: an electrostatic case (red), electromagnetic case with the true physical $\beta=0.68\%$ (blue), with nominal fast ions included (green), and with reduced fast ions (purple) where fast ion density and gradients were reduced by 20\%. Growth rates are shown in the left panel, frequencies in the right panel}
	\label{fig:section4_EM}
\end{figure}

\subsection{Multiscale simulations with simplified parameters}
\label{sec:multiscale}

The importance of EM-stabilization highlighted in the previous section, led to the realisation that it was not feasible (due to the computational expense of finite-$\beta$ simulations, particularly when long time-scale dynamics are necessary for convergence) to carry out multiscale simulations with sufficient physical realism to correspond directly to the experimental setting. Instead, it was decided to proceed with a reduced physics setting with parameters close to, but not exactly corresponding, to the experimental parameters. See table~\ref{tab:multiscaleparamscomparison} for a comparison. The physical $D/e$ mass ratio was applied throughout. The main reduction in computational cost is due to using a single ion species ($Z_\mathrm{eff}=1$). \rlti$=6.25$ was set to match the ion heat flux in a single-scale nonlinear simulation to the experimental power balance value. {\rlte}$=8$ was set to ensure strongly linearly unstable ETG such that the maximum $\gamma/k_y$ on ETG scales was approximately factor 2 of the maximum $\gamma/k_y$ on ITG scales. This ratio suggests that ETG should play a significant role in the subsequent multiscale simulation. The goal is then to validate QuaLiKiz ETG predictions in an expected clearly ETG-relevant case. {\rlne} was set to 2.3. Both QuaLiKiz-2.6.1 and 2.8.2 predict $Q_e$ caused by high-k ETG fluctuations to approximately equal $Q_i$ caused by low-k ITG fluctuations, leading to a $Q_e\approx Q_i$ scenario since ITG $Q_i/Q_e\approx5$ for these parameters. 

\begin{table*}
    \caption{Main dimensionless physical input parameters applied for the {\gene} and GKV multiscale study, compared with the JINTRAC-QuaLiKiz\_2.6.1 base-case values used in the single-scale study in section~\ref{sec:intmodelling}. $\epsilon$ is the local inverse aspect ratio. Dimensional reference parameters (e.g. for collisionality calculations) were $T_{ref}=2.56~keV$, $n_{ref}=4.87\cdot10^{19}~m^{-3}$, $L_{ref}=3.06~m$.}
    \centering
      \begin{tabular}{c|c|c|c|c|c|c|c|c|c}
	    Case & {\rlti} & {\rlte} & {\rlne} & $\hat{s}$ & $q$ & $T_i/T_e$ & $\alpha$ & $Z_\mathrm{eff}$ & $\epsilon$ \\
        \hline
	    Nominal & 9.25 & 8.62 & 2.93 & 1.5 & 2.02 & 1.08 & 0.62 & 1.65 & 0.67 \\
	    Multiscale study & 6.25 & 8 & 2.3 & 1.5 & 2.02 & 1.08 & 0 & 1 & 0.67 \\
	\end{tabular}
	\label{tab:multiscaleparamscomparison}
	\normalsize
\end{table*}

Both {\gene} and the GKV~\cite{watanabe:2005gkv,maeyama:2013} gyrokinetic codes were applied for the study, to increase validation robustness and to benchmark multiscale predictions between the two different codes. The numerical grid resolution settings are shown in table~\ref{tab:multiscaleparams}.  

\begin{table}
    \caption{{\gene} and GKV grid resolution for the multiscale study. $L_x$ is radial box size in reference ion Larmor units. $n_{kx}$ is the number of radial modes. $k_{y(min)}$ is the minimum normalized wavenumber. $n_z$ is the parallel resolution. $n_v$ and $n_w$ are parallel velocity and perpendicular velocity (magnetic moments) resolution respectively}
    \centering
    \small
     \begin{tabular}{c|c|c|c|c|c|c|c}
	    Code & $L_x$ & $n_{kx}$ & $k_{y(min)}$ & $n_{ky}$ & $n_z$ & $n_v$ & $n_w$\\
        \hline
	    {\gene} & 47.5 & 500 & 0.1 & 620 & 32 & 48 & 16 \\
	    GKV   & 49.5 & 340 & 0.08 & 339 & 32 & 48 & 24 \\
	\end{tabular}
	\label{tab:multiscaleparams}
	\normalsize
\end{table}

\begin{figure}[hbt]
	\centering
	\includegraphics[width=1.0\linewidth]{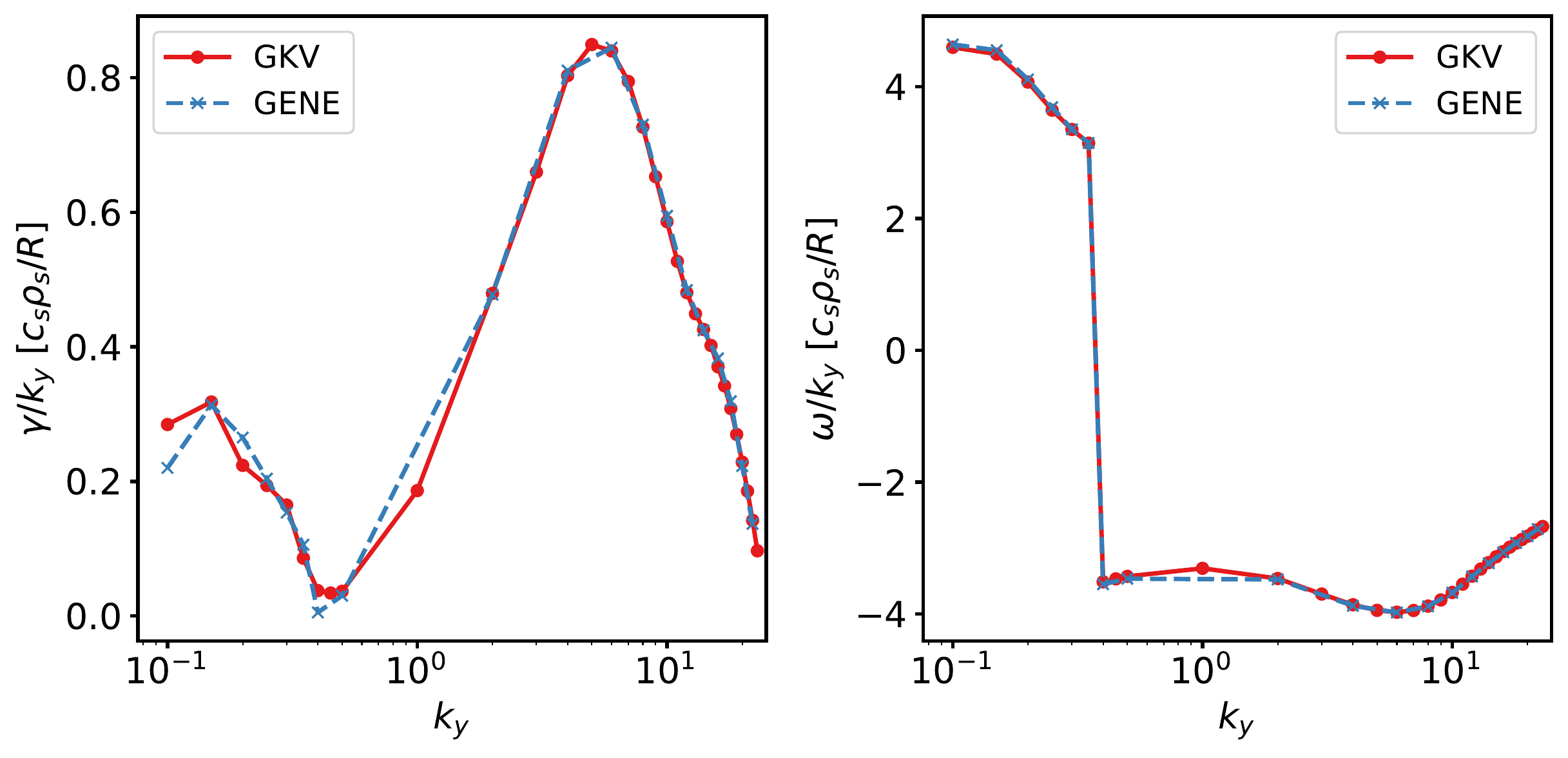}
	\caption{Linear benchmark between GKV and {\gene} for the multiscale parameters for growth rates (left panel) and frequencies (right panel). Both ion and electron scales are shown with a logarithmic x-axis. The growth rates and frequencies are normalized by the wavenumber $k_y$ to maintain similar magnitudes on both ion and electron Larmor radius scales. Modes in the ion diamagnetic direction are defined as having positive frequencies.}
	\label{fig:section4_GKVvsGENElin}
\end{figure}

The linear benchmark between GKV and {\gene} for both ion and electron scales is shown in figure~\ref{fig:section4_GKVvsGENElin}. The codes agree for both spatial-scales, with minor differences observed in ion-scales on the order of 20\%. A like-for-like benchmark was not feasible here since the two codes applied different collision operators: {\gene} a linearized Landau-Bolzmann operator and GKV a Sugama collision operator. This may be the reason for the more extensive relative difference at $k_y=0.4$ which was at the ITG-TEM boundary, and hence extremely sensitive to the collision operator. However, the low growth rate at that $k_y$ means that the impact on the fluxes is negligible.   

\begin{figure}[hbt]
	\centering
	\includegraphics[width=1.0\linewidth]{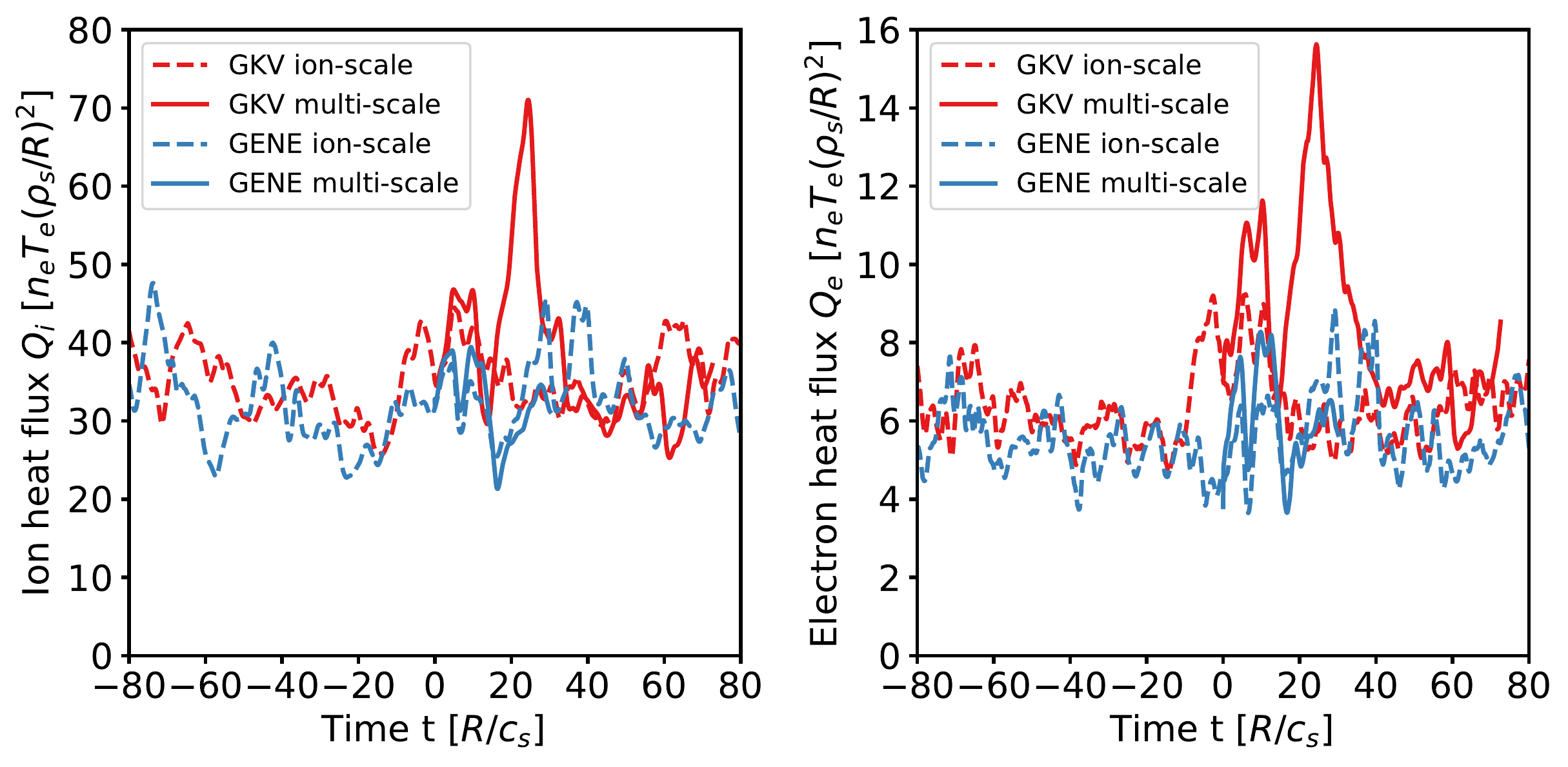} 
	\caption{Temporal evolution of the electron (right panel) and ion (left panel) heat fluxes in GKV (red) and {\gene} (blue) ion-scale (dashed) and multiscale (solid) simulations based on the same physical inputs. The zero in the time axis corresponds to the time when the multiscale simulations are initialized based on the corresponding ion-scale simulation checkpoint. The multiscale GKV and {\gene} simulations do not show a significant addition to the electron heat flux compared to the ion-scale simulations}
	\label{fig:section4_GKVkymax}
\end{figure}

\begin{figure}[hbt]
	\centering
	\includegraphics[width=1.0\linewidth]{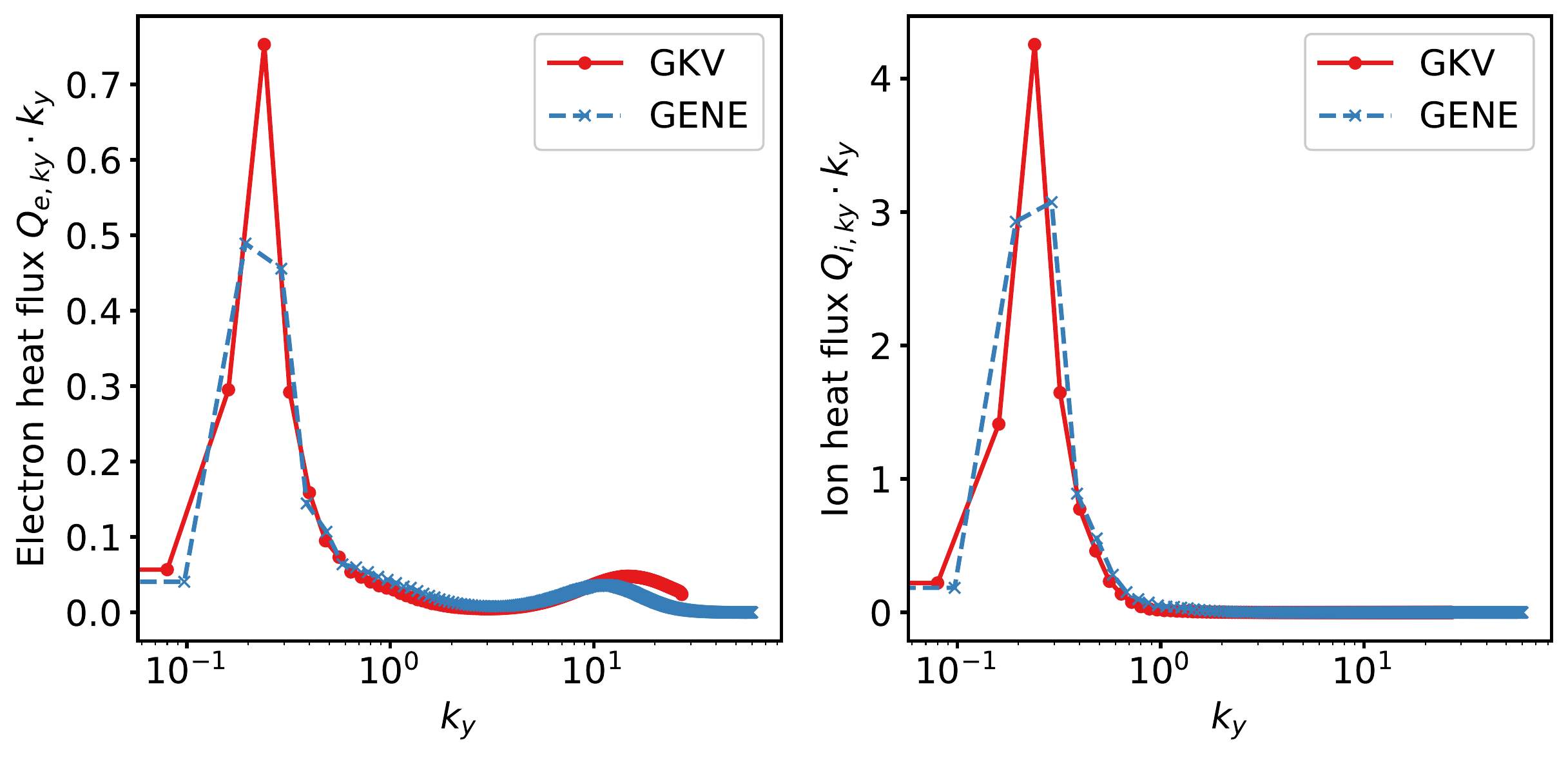}
	\caption{{\gene} and GKV ion (right panel) and electron (left panel) heat flux spectra in a logx plot. The y-axis is $Q_e\cdot k_y$, such that the areas under the curves in the logx plot correspond to the total heat flux. Less than $10\%$ of the electron heat flux is from electron scales. The heat fluxes are in GyroBohm units with $T_\mathrm{ref}=T_e$ and $L_\mathrm{ref}=R_\mathrm{maj}$.}
	\label{fig:section4_multiscale_spectra}
\end{figure}

In the following we discuss the results of the multiscale nonlinear simulations. The central and most striking result of this work, is that in spite of the linear $\left(\gamma/k_y\right)_{\mathrm{maxETG}}\approx2\left(\gamma/k_y\right)_{\mathrm{maxITG}}$, the multiscale simulations did not produce significant ETG heat flux. The results are shown in figures~\ref{fig:section4_GKVkymax} and \ref{fig:section4_multiscale_spectra}. In figure~\ref{fig:section4_GKVkymax}, GKV and {\gene} ion-scale and multiscale simulations are compared. The multiscale simulations were initiated from a checkpoint at a quasi-stationary phase of the corresponding ion-scale nonlinear simulations, and continued for sufficient time to determine whether ETG flux emerges and saturates (see e.g. Fig.1 in Ref.~\cite{maeyama:2017}). This technique allows for faster general convergence by having a good initial condition for the ion-scale $(k_x,k_y)$ in the multiscale simulation. Only an initial transient phase showed an increase in both ion and electron heat fluxes (more evident in the GKV simulations), with the fluxes then converging back to a similar level as the ion-scale simulation. No significant ETG fluxes are predicted. This is further seen in figure~\ref{fig:section4_multiscale_spectra}, where the heat flux spectra from the GKV and {\gene} multiscale simulations are compared. Both codes predict that only $<10\%$ of the electron heat flux arises from electron-scales. A summary of the predicted fluxes by both {\gene} and GKV multiscale simulations is provided in table~\ref{tab:multiscalesummary}.

\begin{table}
    \caption{Summary of nonlinear multiscale {\gene} and GKV ion and electron heat fluxes, averaged over the final 20 time units (in $[R/c_s]$) in their respective simulations. The mean and standard deviation of the fluxes in that time window are provided.}
    \centering
    \small
     \begin{tabular}{c|c|c}
	    Code & $Q_i$ & $Q_e$ \\
        \hline
	    {\gene} & 30.5$\pm$4.5 & 5.9$\pm$1.2   \\
	    GKV   & 32.6$\pm$3.9 & 6.8$\pm$0.8 
	\end{tabular}
	\label{tab:multiscalesummary}
	\normalsize
\end{table}

These results are a counter-example to the classification suggested in Ref.~\cite{creely:2019} (see figure 2 therein). Our case has weak linear TEM and strong linear ETG modes in comparison to the ITG modes, which according to Ref.~\cite{creely:2019} are criteria which should correlate with the importance of multiscale effects. However, we found only weak multiscale effects, hinting at the importance of additional parameters. The results also demonstrate that the rule-of-thumb suggested in Ref.~\cite{howard:2016b} is an insufficient criteria for the onset of significant ETG fluxes. However, generality of this rule was not claimed and due to the relatively small number of multiscale simulations studied, Ref.~\cite{howard:2016b} does not exclude dependence on additional quantities. Furthermore,  the related Eq~\ref{eq:thumbgary} is stated in Ref.~\cite{staebler:2017} as being the non-linear threshold for when ETG linearly exceeds the total suppression due to zonal flow mixing. This does seem to be satisfied in our case, due to the finite (but low) electron-scale bump in the electron heat flux spectrum in figure~\ref{fig:section4_multiscale_spectra}. However, exceeding the non-linear threshold for ETG turbulence does not necessarily correlate to significant ETG flux, which also requires the existence of significant ETG streamers -- elongated structures with $k_x{\ll}k_y$ on electron Larmor radius scales -- which are absent in this case. The $k_x$ spectrum is geometry dependent~\cite{staebler:2021}. Unfortunately there was insufficient computational resources to rerun our cases with more realistic geometry to investigate its impact on ETG streamers. 

We note that while {\rlti} was set to match the power-balance heat flux in the ion-scale simulations, multiscale effects can in principle degrade ion-scale zonal flows and enhance ion-scale flux~\cite{maeyama:2015}. There were insufficient HPC resources to carry out further multiscale simulations at decreased {\rlti} to examine whether in that case significant ETG flux would emerge, with multiscale effects then raising $Q_i$ up to power-balance consistent values. 

To summarize: the multiscale runs carried out showed a lack of significant ETG flux in spite of the strong relative electron to ion scale linear drive. Further investigation of the dependencies which lead to the emergence of significant ETG flux, in cases where the nonlinear ETG threshold has been passed, is out of the scope of this paper. Clearly, there are hidden variables at play which complicate the existence criteria of significant ETG flux beyond simple rules-of-thumb based on the $\gamma/k$ ratios. QuaLiKiz ETG predictions are not validated for this specific parameter set. A more extensive study is necessary to calibrate the QuaLiKiz multiscale rule and quasilinear transport model multiscale saturation rules in general. We note that the cases studied in Refs.~\cite{howard:2016b,creely:2019}, where the rule-of-thumb applies, are in the $Q_e>Q_i$ regime where TEM modes fill the spectral gap between ion and electron scales (see also Figure 5 in Ref.~\cite{staebler:2017}, whereas the cases studied here are in the $Q_i>Q_e$ regime with no spectral gap, as evident in Figure~\ref{fig:section4_GKVvsGENElin}. The ramifications of heat flux ratios on the turbulence regime, spectral gap, and subsequent multiscale impact is left for future work, as is the magnetic geometry impact on streamer formation.

\subsection{Ion-scale simulations with full physics}
\label{sec:miller}

The multiscale results from the previous section suggest that ETG turbulence is not important at $\rho=0.65$ for the discharge studied. This leads to the question of whether ion-scale turbulence in high-fidelity simulations like {\gene} or GKV is sufficient to describe the experimental power balance at the measured kinetic profile gradients for both ion and electron heat flux. We tackle this question when including full-physics on ion-scales, with Miller parameterized geometry~\cite{miller:1998}, inclusion of $E\times B$ shear, EM-effects, and fast ions. Due to limitations of computational time, we limit this study to $\rho=0.65$, and apply {\gene} only. A single effective impurity species was used to save computational cost, maintaining the same impact on main ion dilution and $R/L_{ni}$ as the full set of impurities. 

\begin{figure}[hbt]
	\centering
	\includegraphics[width=0.8\linewidth]{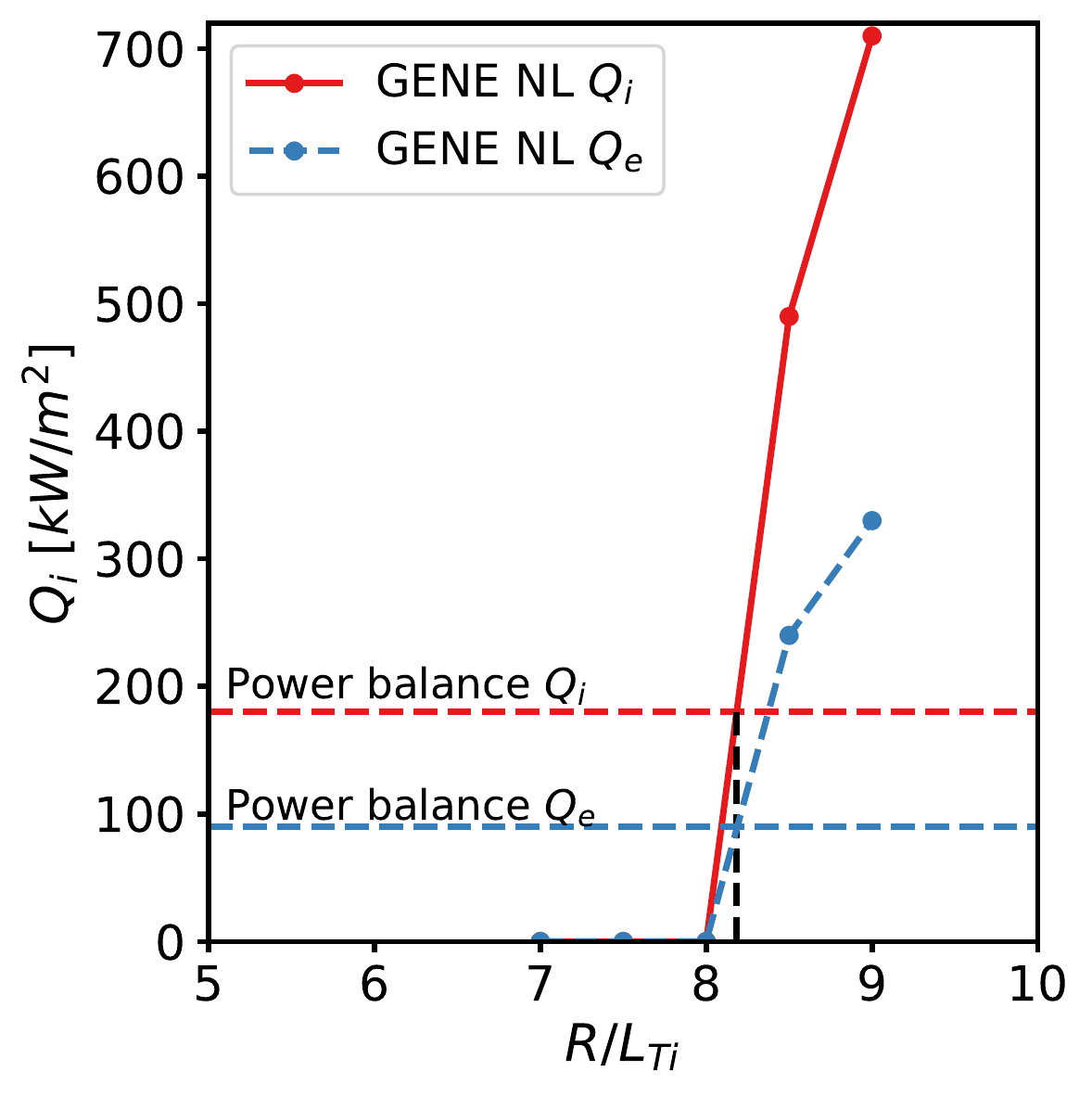}
	\caption{Nonlinear {\gene} ion heat flux (red) and electron heat flux (blue) predictions in a {\rlti} scan at $\rho=0.65$ with both {\rlte} and {\rlne} are reduced by 20\%. The power balance heat fluxes from the integrated modelling run are portrayed by the horizontal dashed lines. Measured {\rlti}$\approx8$}
	\label{fig:section4_NLrltiscan}
\end{figure}

Nonlinear simulations using the nominal measured input parameters (see table~\ref{tab:singlescaleparams}, led to predicted heat fluxes of $Q_i\approx490~kW/m^2$ and $Q_e\approx240~kW/m^2$, significantly above the power balance heat fluxes of $Q_i\approx180~kW/m^2$ and $Q_e\approx90~kW/m^2$. Therefore, an $R/L_{Ti}$ scan was carried out with both $R/L_{Te}$ and $R/L_{ne}$ reduced by 20\%, within their error bars. The results are shown in figure~\ref{fig:section4_NLrltiscan}. The main finding is that experimental power balance is predicted, with the correct experimental heat flux ratio of $Q_i/Q_e\approx2$, within 5\% of the measured {\rlti} when both {\rlte} and {\rlne} are reduced by 20\%. While not the focus of the paper and not shown for brevity, particle transport was overpredicted by a factor $\sim2$ and momentum transport underpredicted by a factor $\sim2$. More extensive uncertainty quantification including propagation of gradient, magnetic equilibrium, fast ion, and impurity uncertainties, and an attempt to simultaneously match all transport channels within these input certainties, was not feasible due to lack of resources. However, the limited uncertainty quantification performed already demonstrates that ion-scale turbulence predictions provides a consistent description of the heat fluxes, and that it is not necessary to invoke ETG turbulence to describe the electron heat flux for this case. 

In the course of this exercise, a number of interesting observations were made regarding the impact of EM-effects at finite-$\beta$. These are described below.

\begin{figure}[hbt]
	\centering
	\includegraphics[width=1.0\linewidth]{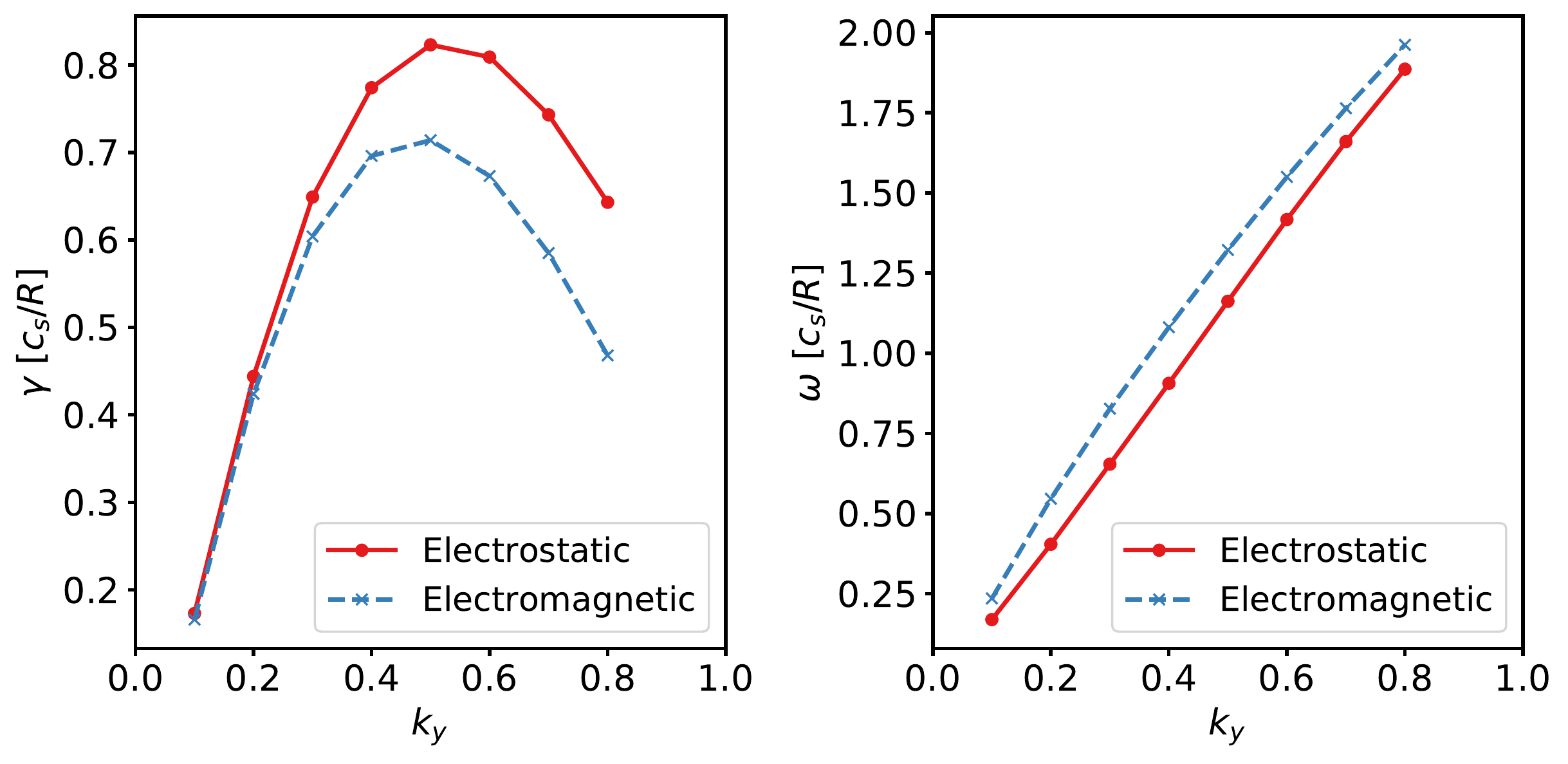}
	\caption{Electrostatic (red curves) vs electromagnetic (blue curves) linear {\gene} simulations with Miller geometry, at nominal {\rlti} and both {\rlte} and {\rlne} reduced by 20\%. No rotation is included. The growth rate spectrum is shown in the left panel, and the frequency spectrum in the right panel.}
	\label{fig:section4_linearmillerEMvsES}
\end{figure}

Firstly, the linear EM impact in Miller geometry for this parameter set is much reduced compared to the analogous {\sa} case. This is seen in figure~\ref{fig:section4_linearmillerEMvsES}, showing a comparison between EM and ES linear calculations for nominal {\rlti} and both {\rlte} and {\rlne} reduced by 20\%. The linear impact of finite-$\beta$ can be compared to the {\sa} case in figure~\ref{fig:section4_EM}.  $\alpha_\mathrm{MHD}$ is kept fixed at the nominal value for both the EM and ES simulations. While in the {\sa} case the growth rate reduction due to EM-effects was significant across the entire spectrum, in the Miller case the impact is much weakened, particularly at the transport driving scales at lower $k_y$. A weak linear EM-stabilization effect at outer radii was also observed in Ref.~\cite{citrin:2015b} with shaped geometry. The comparison between EM-stabilization in shaped and $\hat{s}-\alpha$ geometry was not carried out in previous work. An investigation regarding the physical provenance of the observed differences of EM-stabilization due to geometry is interesting also due to a possibility for performance optimization with shaping actuators, but is out of the scope of this paper. 

\begin{figure}[hbt]
	\centering
	\includegraphics[width=1.0\linewidth]{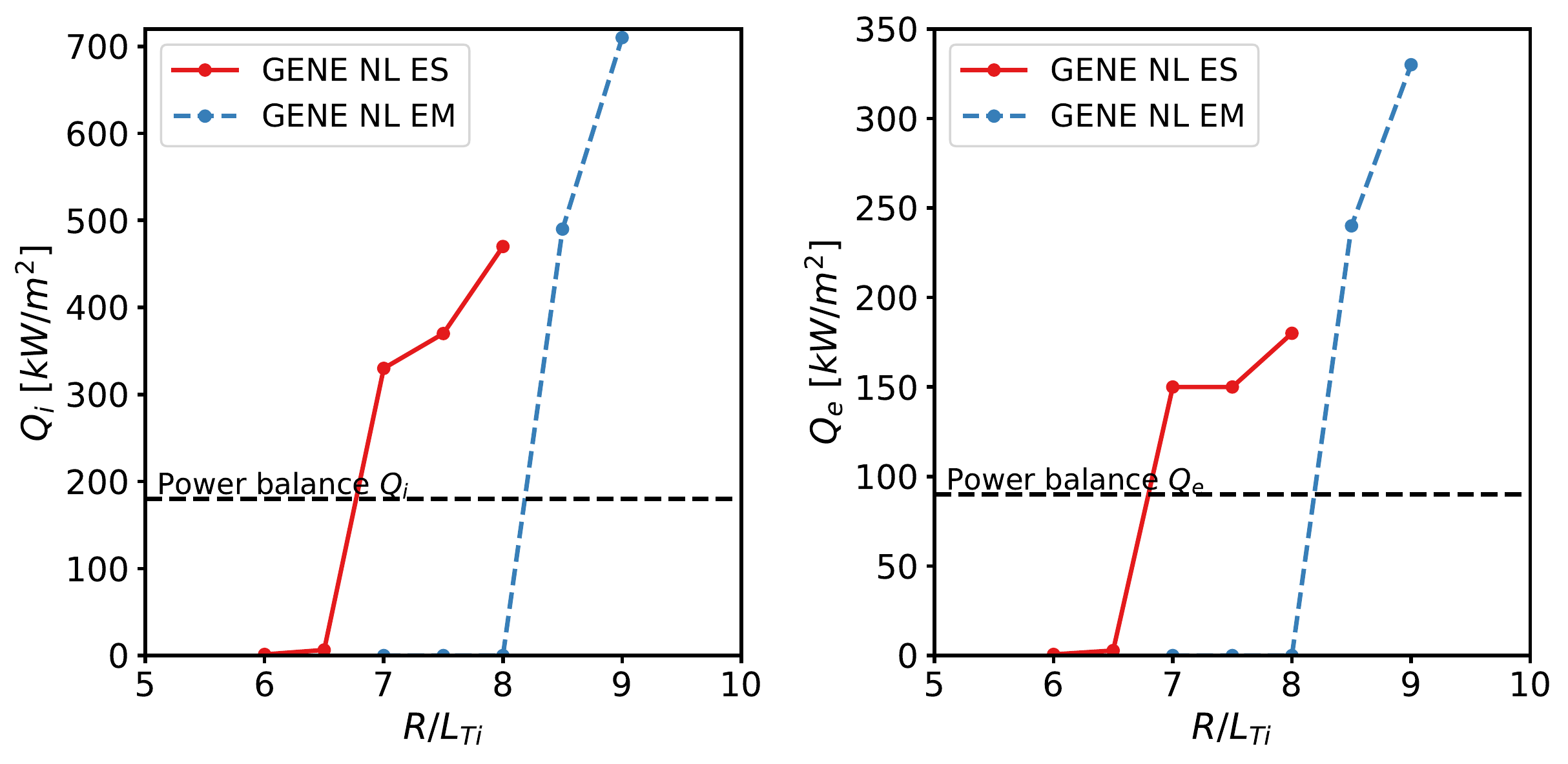}
	\caption{{\rlti} scan from both electrostatic (red) vs electromagnetic (blue) nonlinear {\gene} simulations with Miller geometry. Both {\rlte} and {\rlne} are reduced by 20\% from the nominal measured values (see table~\ref{tab:singlescaleparams}). The power balance heat fluxes from the integrated modelling run is portrayed by the horizontal dashed lines.}
	\label{fig:section4_NL_ES_EM_compare}
\end{figure}

Nevertheless, even though the linear impact is minor, significant EM-stabilization is still observed in the nonlinear {\gene} simulations with Miller geometry. This is seen in figure~\ref{fig:section4_NL_ES_EM_compare}, showing the {\rlti} scan at 20\% reduced {\rlte} and {\rlne}, for both the (nominal) EM simulations as well as an ES scan. For both ES and EM cases, the transport is stiff with power balance reached in the proximity of the nonlinear {\rlti} threshold. However, an {\rlti} nonlinear threshold upshift on the order of $\approx20\%$ is apparent in the EM scan compared to the ES scan. This is suggestive of finite-$\beta$ effects extending the Dimits shift regime~\cite{dimits:2000}, as in Ref.~\cite{pueschel:2010}. The upshift is of the same order of the ad-hoc threshold upshift EM-stabilization rule applied in QuaLiKiz, supporting its use. However, while encouraging, no general statement can yet be made regarding the validity of the QuaLiKiz ad-hoc EM-stabilization rule, as in other studies the primary impact of EM-stabilization was predicted to be a reduction of transport stiffness (slope of $Q_i$ with respect to {\rlti})~\cite{citrin:2015b}. Additional work is necessary to untangle the parameterization and impact of EM-stabilization in reduced models.

\begin{figure}[hbt]
	\centering
	\includegraphics[width=1.0\linewidth]{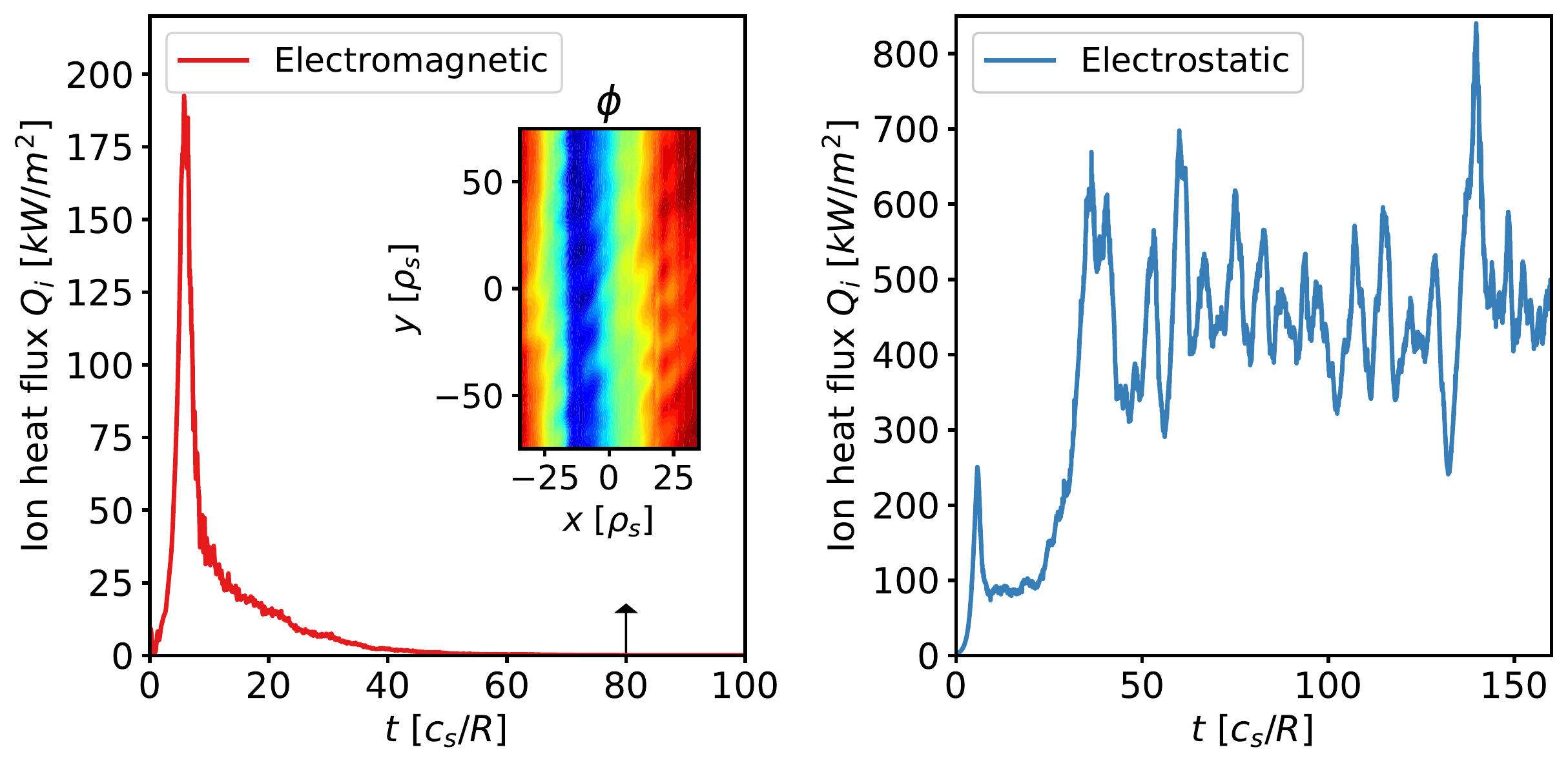}
	\caption{Comparison of electrostatic (right panel) and electromagnetic (left panel) nonlinear {\gene} simulation ion heat flux time series. {\rlti}$=8$ and both {\rlte} and {\rlne} are reduced by 20\% from their nominal measured values. The EM case electrostatic potential contour at $t=80$, at the $z$ position corresponding to the LFS, is inset in the left panel.}
	\label{fig:section4_ESvsEM_nrg}
\end{figure}

The heightened impact of zonal flows on the nonlinear threshold upshift in the electromagnetic simulation is evident in figure~\ref{fig:section4_ESvsEM_nrg}. $Q_i$ time-traces are shown from the {\rlti}$=8$ EM and ES simulations at 20\% reduced {\rlte} and {\rlne}. The electrostatic case is robustly unstable. The electromagnetic case, following the initial linear phase, transitions into a quiescent state dominated by zonal flows, as seen by the electrostatic potential contour map inset, where the flux asymptotically approaches zero. 

Interestingly, while in previous work on gyrokinetic analysis of JET hybrid scenarios, a weak linear EM-stabilization at outer radii was also observed with shaped geometry~\cite{citrin:2015b}, the previous work also showed no significant nonlinear EM-stabilization in contrast to the present study. The most significant difference in parameters between the previous and present work, are the thermal $R/L_{Ti}$ and $R/L_{Te}$ values, which are much higher in the present study, likely owing to the increased heating power. These parameters contribute to the EM-mode drive, perhaps contributing to the stabilization. A detailed study is out of the scope of this work.

\begin{figure}[hbt]
	\centering
	\includegraphics[width=1.0\linewidth]{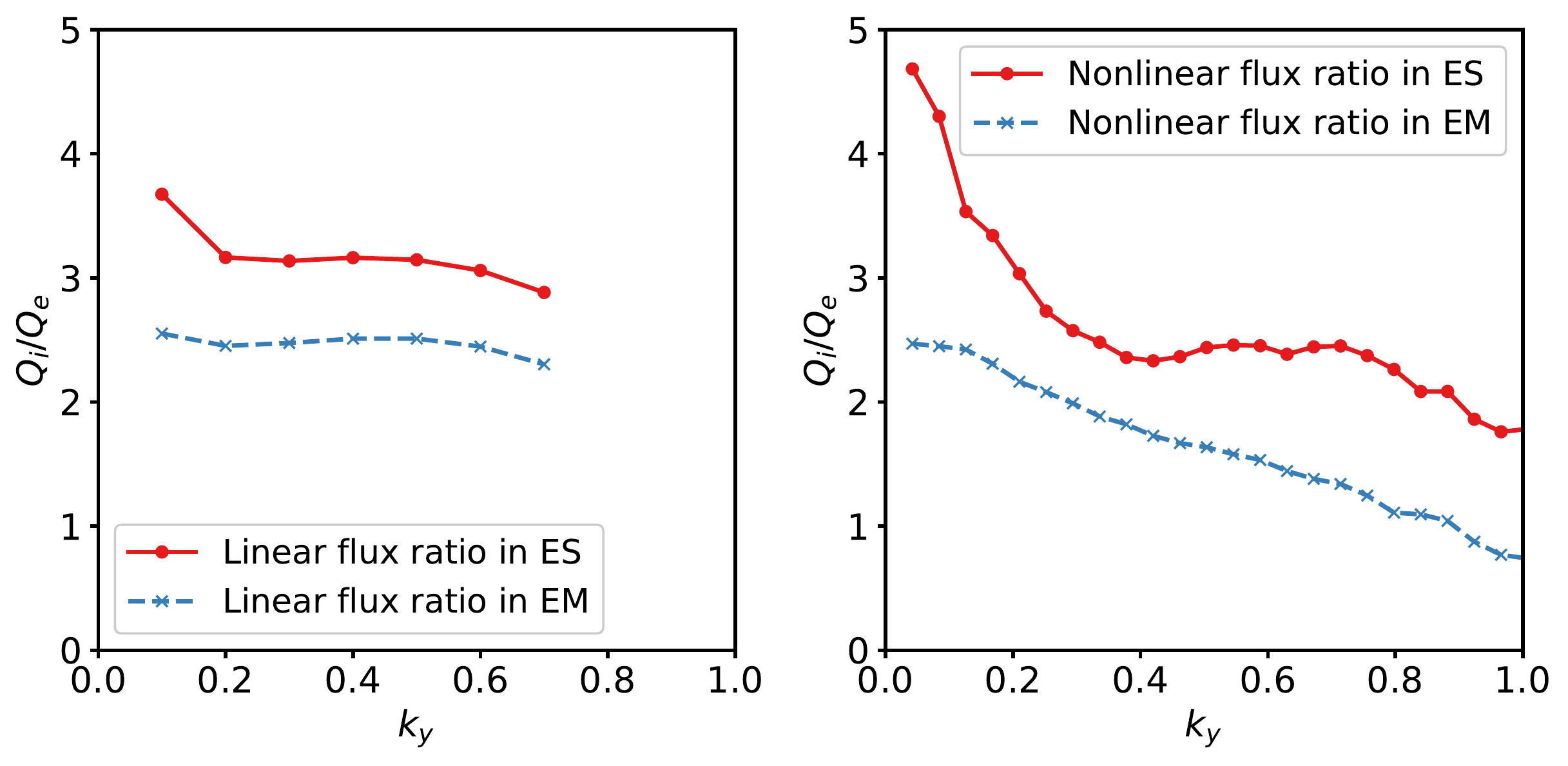}
	\caption{Comparison of ion to electron heat flux ratios in electrostatic and electromagnetic cases from linear {\gene} calculations (left panel) and Fourier decomposition of a nonlinear {\gene} simulation (right panel). {\rlti}$=8.5$, with both {\rlte} and {\rlne} reduced by 20\% from their nominal measured quantities.}
	\label{fig:section4_fluxratiosEMvsES}
\end{figure}

An additional impact of EM-effects was observed on the ion-scale electron heat flux. With finite-$\beta$, the electron heat flux was observed to increase relative to the ion heat flux. The increase was in the electrostatic heat flux, with the electromagnetic heat flux (magnetic flutter) remaining negligible. The electron heat flux increase is likely dominated by linear effects, since a similar relative increase is seen both in linear and nonlinear simulations, as shown in figure~\ref{fig:section4_fluxratiosEMvsES}. Averaging over the ion-scale $k_y$ spectrum, the linear $Q_i/Q_e$ decreased from 3.26 to 2.50 when transitioning from ES to EM simulations. The nonlinear case decreased from 2.86 to 2.01. We conjecture that this effect is related to the mechanism previously predicted to increase electron particle flux in finite-$\beta$ gyrokinetic calculations~\cite{hein:2010}. The next section further explores this conjecture, which contributes towards reaching the experimental power balance of $Q_i/Q_e\approx2$. 

\section{Analytic study of $\beta$-enhanced electron heat flux}
This section derives an analytical model that isolates the impact of finite-$\beta$ on the non-magnetic-flutter component of the electron heat flux. The derivation is strongly based on related work in Ref.~\cite{hein:2010} on the impact of finite-$\beta$ on particle flux. The analytical model provides a prediction that the ion to electron heat flux ratio is decreased in the ITG regime, and potentially increased in the TEM regime. This prediction is validated by dedicated numerical gyrokinetic calculations.

We now proceed with the derivation. The electron heat flux can be divided in trapped and passing components:

\begin{eqnarray}
q_{e} = f_t q_{e,trap} + (1-f_t) q_{e,pass}
\label{elheatfluxtrappass}
\end{eqnarray}

where $q_{e,trap}$ and $q_{e,pass}$ are the respective trapped and passing electron contributions normalized to their population density.
Finite-$\beta$ effects only impact $ q_{e,pass}$ due to the role of parallel velocity in Amp\`{e}re's Law. As such, we concentrate on $q_{e,pass}$.

Closely following the derivation in Ref.~\cite{hein:2010}, the general expression for the quasilinear electron heat flux, neglecting magnetic flutter terms, is:
\small
\begin{eqnarray}
\fl q_{e,pass} \\ = \sum_k \left\langle k_y \rho_s c_s \int d^3v E F_0 J \left[1-2\vpar \Omega_{r,k}+\vpar^2\left(\Omega_{r,k}^2+\Omega_{i,k}^2\right)\right]  
H \right\rangle
\nonumber \\
\fl H=  \\ \Phi^2 \frac{(\gamma_k+\nu)k_y\rho_s [\frac{R}{L_{ne}} +(E-\frac{3}{2}\frac{R}{L_{Te}})]-[\gamma_k(\kpar\vpar+\omega_{d,k})-\omega_{r,k}\nu]}{(\omega_{r,k}+\kpar\vpar+\omega_{d,k})^2+(\gamma_k+\nu)^2} \nonumber
\label{qepassdef}
\end{eqnarray}
\normalsize
where $E$ is normalized (to $T_e$) particle energy, $J$ the Bessel function due to FLR effects, $k$ the respective mode number, $\nu$ the collisionality, $\gamma$ the mode growth rate, $\omega_r$ its real frequency, $\omega_d$ the $\nabla B$ drift frequency, and $\Omega_{r,k},\Omega_{i,k}$ are the real and imaginary part of the quantity $\Omega$ defined as:
\begin{eqnarray}
\Apar=\Omega\Phi
\label{omegadef}
\end{eqnarray}
where $\Apar$ is the fluctuating vector potential, and $\Phi$ the fluctuating electrostatic potential.
The terms proportional to $\Omega$ represent the contribution of $\beta$ to the non-magnetic-flutter component of the electron heat flux. Upon solving the Amp\`{e}re-Maxwell system, one finds that $\Omega$, in the limit of large parallel velocity of passing electrons, becomes:
\begin{eqnarray}
\Omega_k \approx \frac{q\beta(k_y\rho_sR/L_{ne}+\omega_{r,k}+i\gamma)}{(k_\perp\rho_s)^2}
\label{omegardef}
\end{eqnarray}
In the low $\beta$ limit, we neglect the terms proportional to $\Omega^2$. Moreover, we take the limit $\vpar \rightarrow \infty$ (because of the mass ratio appearing when normalizing to $c_s$). The leading order term of the $\beta$-dependent $E{\times}B$ electron heat flux is then:
\begin{eqnarray}
\fl q_{e,pass,\beta} \\ \approx \sum_k \left\langle k_y \rho_s c_s \int d^3v E F_0 J \left[-2\vpar \Omega_{r,k}\right]  \Phi^2 \frac{-[\gamma_k\kpar\vpar]}{(\kpar\vpar)^2} \right\rangle \nonumber
\label{qepassinfty}
\end{eqnarray}
that is:
\begin{eqnarray}
\fl q_{e,pass,\beta} \\ \approx \beta \sum_k \left\langle 2qk_y \rho_s c_s  \frac{k_y\rho_sR/L_{ne}+\omega_{r,k}}{(k_\perp\rho_s)^2}  \Phi^2 \frac{\gamma_k}{\kpar} \int d^3v E F_0 J   \right\rangle
\label{qepassinfty2} \nonumber
\end{eqnarray}
For ITG (positive $\omega$), this gives an outward heat flux contribution, proportional to the plasma energy and $\beta$. For TEM (negative $\omega$), particularly at non-extreme density peaking scenarios, this can lead to an inward heat flux contribution.

\begin{figure}[hbt]
	\centering
	\includegraphics[width=0.8\linewidth]{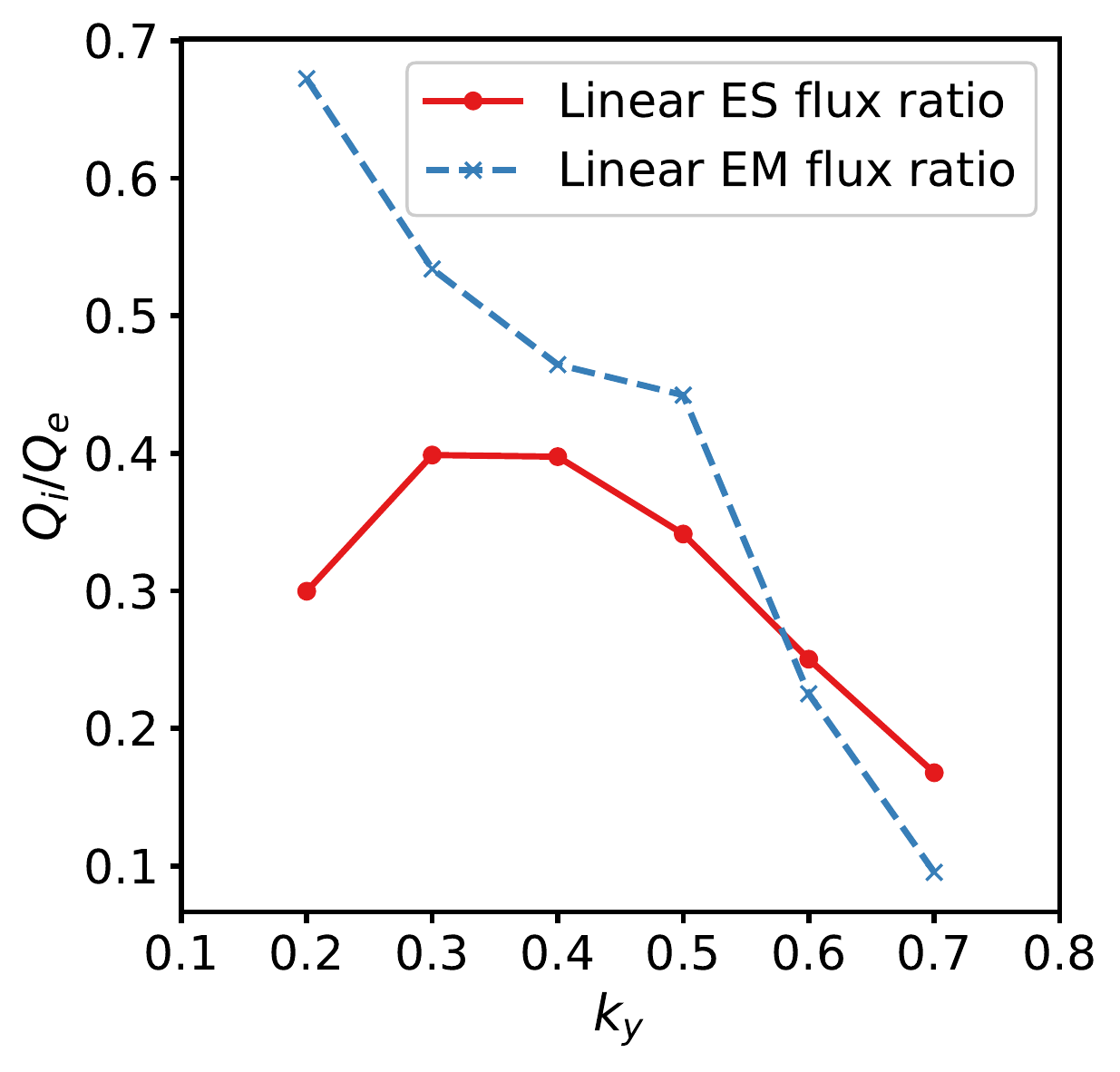}
	\caption{Comparison of ion to electron heat flux ratios in electrostatic and electromagnetic cases from linear {\gene} calculations corresponding to a TEM case. Nominal measured parameters are used apart from {\rlti}$=3$, such that TEMs dominate the spectrum.}
	\label{fig:section4_fluxratiosEMvsES_TEM}
\end{figure}

The analytic prediction of the differing sign of $\beta$-induced $E\times B$ electron heat flux for ITG modes vs TEMs is borne out by linear gyrokinetic simulations for the case studied here. Figure~\ref{fig:section4_fluxratiosEMvsES_TEM} shows the heat flux ratio for a linear spectrum, corresponding to the nominal measured inputs from table~\ref{tab:singlescaleparams}, with the modification of {\rlti}$=3$, such that TEMs are the dominant unstable modes. For the majority of the spectrum, the ES-case $Q_i/Q_e$ flux ratio is now below the EM-case $Q_i/Q_e$ flux ratio, as opposed to the ITG cases previously shown. 

\section{Conclusions}
\label{sec:conclusions}
The primary aim of this work was to validate predictions that in high $T_i/T_e$ hybrid H-mode scenarios on JET, ETG turbulence plays an important role in clamping $T_e$, leading to increased $T_i/T_e$ and improved ion thermal confinement at heavier isotopes (anti-Gyrobohm isotope confinement scaling), enabled due to a reduction in ion-electron heat exchange at heavier isotope. These previous results were obtained in integrated modelling with a version of QuaLiKiz in which the collision operator was incorrectly overly stabilizing TEMs~\cite{casson:2020}. Following improvements of the QuaLiKiz collision operator~\cite{stephens:2021b}, these JET hybrid H-mode predictions were revisited both with the improved QuaLiKiz as well as high-fidelity nonlinear gyrokinetic single-scale and multiscale modelling. Understanding the role of ETG turbulence in these high-performance scenarios, and its impact on isotope confinement scaling, is of great importance towards physics interpretation of T and DT JET hybrid H-mode scenarios, and in validating reduced transport models applied for extrapolations towards ITER and reactors. 

While the outlined mechanism is valid, the results in this paper invalidate the specific previous prediction by showing reduced evidence of the importance of ETG turbulence in this regime. Point-by-point conclusions are as follows. As a caveat, we stress that all gradient-driven analysis involving higher-fidelity {\gene} and GKV modelling are restricted to $\rho=0.65$ of the analyzed JET discharge. 

\begin{itemize}
    \item Integrated modelling of a similar JET hybrid H-mode scenario as in Ref.~\cite{casson:2020} with QuaLiKiz-2.8.2 (improved collision operator) predicts a reduced significance of ETG turbulence and its corresponding anti-GyroBohm scaling effect. Unstable ETG is limited to the $\rho<0.4$ region with QuaLiKiz-2.8.2, instead of broadly as predicted by the integrated modelling with QuaLiKiz-2.6.1.
    \item {\gene} modelling validates QuaLiKiz-2.8.2 ion-scale predictions in the power balance relevant regime, and invalidates QuaLiKiz-2.6.1 predictions which had too high $Q_i/Q_e$ on ion scales due to over-suppression of TEM. 
    \item {\gene} and GKV multiscale nonlinear simulations in an expected ETG-relevant regime with $\left(\gamma/k_y\right)_{\mathrm{maxETG}}>\left(\gamma/k_y\right)_{\mathrm{maxITG}}$, did not predict significant ETG flux. This suggests that additional hidden variables impact the emergence of multiscale ETG flux. Simple rules-of-thumb based on linear drive ratios may indicate the emergence of finite ETG flux above nonlinear thresholds, but not its significance, which depends on additional quantities impacting streamer formation and multiscale interactions.
    \item {\gene} nonlinear ion-scale modelling with realistic geometry, rotation and EM effects at $\rho=0.65$ was sufficient to recover the experimental power balance heat flux ratio without the need to invoke ETG turbulence. The power balance fluxes themselves were reproduced with {\rlti} within 5\%, and both {\rlte} and {\rlne} values within $\approx20\%$, of the measured values. 
\end{itemize}

The {\gene} nonlinear ion-scale analysis triggered further investigations regarding the impact of EM-effects on turbulence. These secondary conclusions are as follows:

\begin{itemize}
    \item A significant difference in linear EM-stabilization of ITG modes was found between simulations with {\sa} and more realistic shaped Miller geometry. The linear EM-stabilization impact was reduced with Miller geometry. However, with Miller geometry a significantly enhanced Dimits shift regime was observed in nonlinear simulations, leading to a 20\% increase in {\rlti}, key to reaching power balance fluxes within experimental gradient error bars. The nature and magnitude of the threshold upshift supports the ad-hoc EM-stabilization model in QuaLiKiz.
    \item An increase in the $E \times B$ component of the electron heat flux relative to the ion heat flux was observed in the finite-$\beta$ {\gene} linear and nonlinear simulations. This increase contributes to predicting the inferred heat flux ratio. The effect is related to previous predictions in Ref.~\cite{hein:2010} of the $\beta$ impact on electron particle transport. An analytical prediction of opposite (increased $Q_i/Q_e$ with finite-$\beta$) behaviour for TEM-driven heat flux was validated by gyrokinetic calculations.
\end{itemize}

In general, this work has further illustrated the difficulties of carrying out a multiscale validation effort with realistic parameters. EM effects, impurities, magnetic geometry, and rotation all play key roles in setting the turbulence level. EM effects and impurities increase the required computation time such that extensive validation efforts with uncertainty quantification and numerical convergence checking is barely feasible with current resources. Nevertheless, these inputs are necessary for physical consistency and future efforts should push the envelope in this direction. The inclusion of rotation complicates linking the turbulence predictions with inputs corresponding to power-balance fluxes, to the underlying linear instability drive (e.g. the relative ITG and ETG $\gamma/k$ ratios) due to the time-dependence induced on the linear modes. This further complicates the utility of linear-based metrics for estimating the onset of multiscale effects.

Finally, we stress that while the specific prediction in Ref.~\cite{casson:2020} was not validated by this work, the basic mechanism of ETG-induced anti-GyroBohm scaling is valid. We cannot rule out the emergence of this effect in discharges with further increased $T_i/T_e$, such as the high performance scenarios recently developed at JET~\cite{garcia:2021arxiv}. 

\section{Acknowledgements} \label{Acknowledgements}
This work has been carried out within the framework of the EUROfusion Consortium, funded by the European Union via the Euratom Research and Training Programme (Grant Agreement No 101052200 — EUROfusion). Views and opinions expressed are however those of the author(s) only and do not necessarily reflect those of the European Union or the European Commission. Neither the European Union nor the European Commission can be held responsible for them. The numerical simulations were carried out on the Marconi cluster at CINECA and the JFRS-1 cluster at IFERC-CSC. The authors thank K.L van de Plassche for the QuaLiKiz-pythontools and plotting scripts.

\printbibliography

@STRING{nf	= "Nucl. Fusion" }

@STRING{pop	= "Phys. Plasmas" }

@STRING{prl	= "Phys. Rev. Lett." }

@article{staebler:2021,
  title={Geometry dependence of the fluctuation intensity in gyrokinetic turbulence},
  author={Staebler, Gary M and Candy, Jeffrey and Belli, Emily A and Kinsey, Jon E and Bonanomi, N and Patel, Bhavin},
  journal={Plasma Physics and Controlled Fusion},
  volume={63},
  number={1},
  pages={015013},
  year={2021},
  publisher={IOP Publishing}
}

@article{appel:2018,
  title={Equilibrium reconstruction in an iron core tokamak using a deterministic magnetisation model},
  author={Appel, LC and Lupelli, I and Contributors, JET},
  journal={Computer Physics Communications},
  volume={223},
  pages={1--17},
  year={2018},
  publisher={Elsevier}
}

@article{belli2020,
  title={Reversal of Simple Hydrogenic Isotope Scaling Laws in Tokamak Edge Turbulence},
  author={Belli, EA and Candy, J and Waltz, RE},
  journal={Physical Review Letters},
  volume={125},
  number={1},
  pages={015001},
  year={2020},
  publisher={APS}
}

@article{garcia:2021arxiv,
  title={{New plasma regimes with small ELMs and high confinement at the Joint European Torus}},
  author={Garcia, J and de la Luna, E and Sertoli, M and Casson, F and Mazzi, S and Stancar, Z and Szepesi, G and Frigione, D and Garzotti, L and Rimini, F and others},
  journal={arXiv preprint arXiv:2103.02679},
  year={2021}
}

@article{bonanomi:2018,
  title={{Impact of electron-scale turbulence and multi-scale interactions in the JET tokamak}},
  author={Bonanomi, N and Mantica, P and Citrin, J and Goerler, T and Teaca, B and Contributors, JET},
  journal={Nuclear Fusion},
  volume={58},
  number={12},
  pages={124003},
  year={2018},
  publisher={IOP Publishing}
}

@article{bourdelle:2015,
  title={{Core turbulent transport in tokamak plasmas: bridging theory and experiment with QuaLiKiz}},
  author={Bourdelle, C and Citrin, J and Baiocchi, B and Casati, A and Cottier, P and Garbet, X and Imbeaux, F and Contributors, JET},
  journal={Plasma Physics and Controlled Fusion},
  volume={58},
  number={1},
  pages={014036},
  year={2015},
  publisher={IOP Publishing}
}

@article{brambilla:2002,
  title={Quasi-local wave equations in toroidal geometry with applications to fast wave propagation and absorption at high harmonics of the ion cyclotron frequency},
  author={Brambilla, Marco},
  journal={Plasma physics and controlled fusion},
  volume={44},
  number={11},
  pages={2423},
  year={2002},
  publisher={IOP Publishing}
}

@misc{	  breslau:2018,
  author	= {J. Breslau and others},
  title		= {TRANSP website: https://transp.pppl.gov},
  institution	= {Princeton Plasma Phys. Lab., PPPL-4088},
  year		= {},
  month		= {},
  doi       ={https://doi.org/10.11578/dc.20180627.4}
}

@article{casson:2020,
  title={{Predictive multi-channel flux-driven modelling to optimise ICRH tungsten control and fusion performance in JET}},
  author={Casson, Francis J and Patten, Hamish and Bourdelle, Clarisse and Breton, Sarah and Citrin, Jonathan and Koechl, Florian and Sertoli, Marco and Angioni, Clemente and Baranov, Yuriy and Bilato, Roberto and others},
  journal={Nuclear Fusion},
  year={2020},
  publisher={IOP Publishing}
}

@TechReport{	  cenacchi:1988,
  title		= {{JETTO: A free boundary plasma transport code}},
  author	= {Cenacchi, G. and Taroni, A.},
  institution	= {ENEA, Rome (Italy)},
  year		= {1988}
}

@article{challis:2015,
  title={{Improved confinement in JET high beta plasmas with an ITER-like wall}},
  author={Challis, Clive D and Garcia, J and Beurskens, M and Buratti, P and Delabie, E and Drewelow, P and Frassinetti, Lorenzo and Giroud, C and Hawkes, N and Hobirk, J and others},
  journal={Nuclear Fusion},
  volume={55},
  number={5},
  pages={053031},
  year={2015},
  publisher={IOP Publishing}
}

@article{citrin:2015b,
  title={Electromagnetic stabilization of tokamak microturbulence in a high-beta regime},
  author={Citrin, J and Garcia, J and G{\"o}rler, T and Jenko, F and Mantica, P and Told, D and Bourdelle, C and Hatch, DR and Hogeweij, GMD and Johnson, Thomas and others},
  journal={Plasma Physics and Controlled Fusion},
  volume={57},
  number={1},
  pages={014032},
  year={2014},
  publisher={IOP Publishing}
}

@article{citrin:2017,
  title={{Tractable flux-driven temperature, density, and rotation profile evolution with the quasilinear gyrokinetic transport model QuaLiKiz}},
  author={J. Citrin and C. Bourdelle and F.J. Casson and C. Angioni and N. Bonanomi and Y. Camenen and X. Garbet and L. Garzotti and T. G{\"o}rler and O. G{\"u}rcan and F. Koechl and F. Imbeaux and O. Linder and K.L. van de Plassche and P. Strand and G. Szepesi JET contributors},
  journal={Plasma Physics and Controlled Fusion},
  volume={59},
  number={12},
  pages={124005},
  year={2017},
  publisher={IOP Publishing}
}

@article{colyer:2017,
  title={{Collisionality scaling of the electron heat flux in ETG turbulence}},
  author={Colyer, GJ and Schekochihin, AA and Parra, FI and Roach, CM and Barnes, MA and Ghim, YC and Dorland, W},
  journal={Plasma Physics and Controlled Fusion},
  volume={59},
  number={5},
  pages={055002},
  year={2017},
  publisher={IOP Publishing}
}

@Article{cottier:2013,
  title={Angular momentum transport modeling: achievements of a gyrokinetic quasi-linear approach},
  author={Cottier, P and Bourdelle, C and Camenen, Y and G{\"u}rcan, {\"O} D and Casson, FJ and Garbet, Xavier and Hennequin, P and Tala, Tuomas},
  journal={Plasma Physics and Controlled Fusion},
  volume={56},
  number={1},
  pages={015011},
  year={2013},
  publisher={IOP Publishing}
}

@Article{    creely:2019,
  title={Criteria for the importance of multi-scale interactions in turbulent transport simulations},
  author={A J Creely and P Rodriguez-Fernandez and G D Conway and S J Freethy and N T Howard and A E White and the ASDEX Upgrade Team},
  journal={Plasma Physics and Controlled Fusion},
  volume={61},
  pages={085022},
  year={2019},
  publisher={IOP Publishing}
}

@Article{	  dimits:2000,
  author	= {A.M. Dimits and G. Bateman and M.A. Beer and B.I. Cohen
		  and W. Dorland and G.W. Hammett and C. Kim and J.E. Kinsey
		  and M. Kotschenreuther and A.H. Kritz and L.L. Lao and J.
		  Mandrekas and W.M. Nevins and S.E. Parker and A.J. Redd and
		  D.E. Shumaker and R. Sydora and J. Weiland},
  title		= {Comparisons and physics basis of tokamak transport models
		  and turbulence simulations},
  journal	= pop,
  volume	= 7,
  pages		= 969,
  year		= 2000
}

@article{disiena:2021,
  title={Nonlinear electromagnetic interplay between fast ions and ion-temperature-gradient plasma turbulence},
  author={Di Siena, A and G{\"o}rler, T and Poli, E and Navarro, A Ba{\~n}{\'o}n and Biancalani, A and Bilato, R and Bonanomi, N and Novikau, I and Vannini, F and Jenko, F},
  journal={Journal of Plasma Physics},
  volume={87},
  number={2},
  year={2021},
  publisher={Cambridge University Press}
}

@article{doerk:2015,
  title={{Electromagnetic effects on turbulent transport in high-performance ASDEX Upgrade discharges}},
  author={Doerk, H and Dunne, M and Jenko, F and Ryter, F and Schneider, PA and Wolfrum, E and ASDEX Upgrade Team},
  journal={Physics of Plasmas},
  volume={22},
  number={4},
  pages={042503},
  year={2015},
  publisher={AIP Publishing LLC}
}

@Article{	  dorland:2000,
  author	= {W. Dorland and F. Jenko and M. Kotschenreuther and B.N.
		  Rogers},
  title		= {Electron temperature gradient turbulence},
  journal	= prl,
  volume	= 85,
  pages		= 5579,
  year		= 2000
}

@Article{	  garbet:2010,
  title		= {Gyrokinetic simulations of turbulent transport},
  author	= {Garbet, X and Idomura, Y and Villard, L and Watanabe, TH},
  journal	= nf,
  year		= {2010},
  number	= {4},
  pages		= {043002},
  volume	= {50},
  publisher	= {IOP Publishing}
}

@article{    garcia:2015,
  title={Key impact of finite-beta and fast ions in core and edge tokamak regions for the transition to advanced scenarios},
  author={Garcia, J and Challis, C and Citrin, J and Doerk, H and Giruzzi, G and G{\"o}rler, T and Jenko, F and Maget, P and Contributors, JET},
  journal={Nuclear Fusion},
  volume={55},
  number={5},
  pages={053007},
  year={2015},
  publisher={IOP Publishing}
}

@article{goldston:1981,
  title={New techniques for calculating heat and particle source rates due to neutral beam injection in axisymmetric tokamaks},
  author={Goldston, Robert James and McCune, DC and Towner, HH and Davis, SL and Hawryluk, RJ and Schmidt, GL},
  journal={Journal of computational physics},
  volume={43},
  number={1},
  pages={61--78},
  year={1981},
  publisher={Elsevier}
}

@Article{    grierson:2018,
author = {Grierson,B. A.  and Staebler,G. M.  and Solomon,W. M.  and McKee,G. R.  and Holland,C.  and Austin,M.  and Marinoni,A.  and Schmitz,L.  and Pinsker,R. I. },
title = {{Multi-scale transport in the DIII-D ITER baseline scenario with direct electron heating and projection to ITER}},
journal = {Physics of Plasmas},
volume = {25},
number = {2},
pages = {022509},
year = {2018},
doi = {10.1063/1.5011387}
}

@article{ho:2019,
  title={Application of Gaussian process regression to plasma turbulent transport model validation via integrated modelling},
  author={Ho, A and Citrin, J and Auriemma, F and Bourdelle, C and Casson, FJ and Kim, Hyun-Tae and Manas, P and Szepesi, G and Weisen, H and Contributors, JET},
  journal={Nuclear Fusion},
  volume={59},
  number={5},
  pages={056007},
  year={2019},
  publisher={IOP Publishing}
}

@article{hobirk:2012,
  title={{Improved confinement in JET hybrid discharges}},
  author={Hobirk, J and Imbeaux, F and Crisanti, F and Buratti, P and Challis, CD and Joffrin, E and Alper, B and Andrew, Y and Beaumont, P and Beurskens, M and others},
  journal={Plasma Physics and Controlled Fusion},
  volume={54},
  number={9},
  pages={095001},
  year={2012},
  publisher={IOP Publishing}
}

@article{holland:2017,
  title={{Gyrokinetic predictions of multiscale transport in a DIII-D ITER baseline discharge}},
  author={Holland, C and Howard, NT and Grierson, BA},
  journal={Nuclear Fusion},
  volume={57},
  number={6},
  pages={066043},
  year={2017},
  publisher={IOP Publishing}
}

@article{      houlberg:1997,
author = {Houlberg,W. A.  and Shaing,K. C.  and Hirshman,S. P.  and Zarnstorff,M. C. },
title = {Bootstrap current and neoclassical transport in tokamaks of arbitrary collisionality and aspect ratio},
journal = {Physics of Plasmas},
volume = {4},
number = {9},
pages = {3230-3242},
year = {1997},
doi = {10.1063/1.872465},
}

@Article{	  howard:2014c,
  author		= {Howard, N. T. and White, A. E. and Greenwald, M. and
		  	Holland, C. and Candy, J.},
  title			= {{Multi-scale gyrokinetic simulation of Alcator C-Mod
		  	tokamak discharges}},
  journal		= pop,
  volume		= {21},
  pages		= {032308},
  year		= {201},
  doi			= {http://dx.doi.org/10.1063/1.4869078}
}

@Article{	  howard:2016a,
  author	= {N.T. Howard and C. Holland and A.E. White and M. Greenwald
		  and J. Candy},
  title		= {Multi-scale gyrokinetic simulation of tokamak plasmas:
		  enhanced heat loss due to cross-scale coupling of plasma
		  turbulence},
  journal	= nf,
  volume	= 56,
  number	= 1,
  pages		= 014004,
  year		= 2016
}

@Article{	  howard:2016b,
  author	= "Howard, N. T. and Holland, C. and White, A. E. and
		  Greenwald, M. and Candy, J. and Creely, A. J.",
  title		= "Multi-scale gyrokinetic simulations: Comparison with
		  experiment and implications for predicting turbulence and
		  transport",
  journal	= pop,
  year		= "2016",
  volume	= "23",
  pages		= 056109,
  doi		= "http://dx.doi.org/10.1063/1.4946028"
}

@Article{	  jenko:2000b,
  author	= {F. Jenko and W. Dorland and M. Kotschenreuther and B.N.
		  Rogers},
  title		= {Electron temperature gradient driven turbulence},
  journal	= pop,
  volume	= 7,
  pages		= 1904,
  year		= 2000
}

@Article{	  jenko:2001b,
  author	= {F. Jenko and W. Dorland and G.W. Hammett},
  title		= {Critical gradient formula for toroidal electron
		  temperature gradient modes},
  journal	= pop,
  volume	= 8,
  pages		= 4096,
  year		= 2001
}

@Article{	  jenko:2002,
  author	= {F. Jenko and W. Dorland},
  title		= {Prediction of Significant Tokamak Turbulence at Electron
		  Gyroradius Scales},
  journal	= prl,
  volume	= 89,
  pages		= {225001-1},
  year		= 2002
}

@article{joffrin:2005,
  title={{The ‘hybrid’ scenario in JET: Towards its validation for ITER}},
  author={Joffrin, E and Sips, ACC and Artaud, JF and Becoulet, A and Bertalot, L and Budny, R and Buratti, P and Belo, P and Challis, CD and Crisanti, F and others},
  journal={Nuclear fusion},
  volume={45},
  number={7},
  pages={626},
  year={2005},
  publisher={IOP Publishing}
}

@article{kumar:2021,
  title={{Turbulent transport driven by kinetic ballooning modes in the inner core of JET hybrid H-modes}},
  author={Kumar, Neeraj and Camenen, Yann and Benkadda, Sadruddin and Bourdelle, Clarisse and Loarte, Alberto and Polevoi, Alexei R and Widmer, Fabien and others},
  journal={Nuclear Fusion},
  volume={61},
  number={3},
  pages={036005},
  year={2021},
  publisher={IOP Publishing}
}

@Article{	  lao:1985,
  author	= {L.L. Lao and H. St. John and R.D. Stambaugh and A.G.
		  Kellman and W. Pfeiffer},
  title		= {Reconstruction of current profile parameters and plasma
		  shapes in tokamaks},
  journal	= nf,
  volume	= 25,
  pages		= 1611,
  year		= 1985
}

@article{maeyama:2013,
  title={{Numerical techniques for parallel dynamics in electromagnetic gyrokinetic Vlasov simulations}},
  author={Maeyama, Shinya and Ishizawa, Akihiro and Watanabe, T-H and Nakajima, Noriyoshi and Tsuji-Iio, Shunji and Tsutsui, Hiroaki},
  journal={Computer Physics Communications},
  volume={184},
  number={11},
  pages={2462--2473},
  year={2013},
  publisher={Elsevier}
}

@article{maeyama:2015,
  title={Cross-scale interactions between electron and ion scale turbulence in a tokamak plasma},
  author={Maeyama, S and Idomura, Y and Watanabe, T-H and Nakata, M and Yagi, M and Miyato, N and Ishizawa, A and Nunami, M},
  journal={Physical review letters},
  volume={114},
  number={25},
  pages={255002},
  year={2015},
  publisher={APS}
}

@article{maeyama:2017,
  title={Suppression of ion-scale microtearing modes by electron-scale turbulence via cross-scale nonlinear interactions in tokamak plasmas},
  author={Maeyama, S and Watanabe, T-H and Ishizawa, A},
  journal={Physical Review Letters},
  volume={119},
  number={19},
  pages={195002},
  year={2017},
  publisher={APS}
}

@article{mantica:2021,
  title={{The role of electron-scale turbulence in the JET tokamak: experiments and modelling}},
  author={Mantica, P and Bonanomi, N and Mariani, A and Carvalho, P and Delabie, E and Garcia, J and Hawkes, N and Johnson, T and Keeling, D and Sertoli, M and others},
  journal={Nuclear Fusion},
  volume={61},
  number={9},
  pages={096014},
  year={2021},
  publisher={IOP Publishing}
}

@article{mariani:2019,
  title={{Investigation of the role of electron temperature gradient modes in electron heat transport in TCV plasmas}},
  author={Mariani, A and Mantica, P and Brunner, S and Fontana, M and Karpushov, A and Marini, C and Porte, L and Sauter, O and TCV Team and EUROfusion MST1 Team and others},
  journal={Nuclear Fusion},
  volume={59},
  number={12},
  pages={126017},
  year={2019},
  publisher={IOP Publishing}
}

@article{mariani:2021,
  title={{Benchmark of quasi-linear models against gyrokinetic single scale simulations in deuterium and tritium plasmas for a JET high beta hybrid discharge}},
  author={Mariani, Alberto and Mantica, Paola and Casiraghi, Irene and Citrin, Jonathan and G{\"o}rler, Tobias and Staebler, Gary M and EUROfusion, JET},
  journal={Nuclear Fusion},
  volume={61},
  number={6},
  pages={066032},
  year={2021},
  publisher={IOP Publishing}
}

@Article{  miller:1998,
  author	= {R.L. Miller and M.S. Chu and J.M. Greene and Y.R. Lin-liu and R.E. Waltz},
  title		= {{Non-circular, finite aspect ratio, local equilibrium model}},  
  journal	= {pop},
  volume	= {5},
  pages		= {973},
  year		= {1998}
}

@article{poli:2018,
  title={{Integrated Tokamak modeling: When physics informs engineering and research planning}},
  author={Poli, Francesca Maria},
  journal={Physics of Plasmas},
  volume={25},
  number={5},
  pages={055602},
  year={2018},
  publisher={AIP Publishing LLC}
}

@article{pueschel:2010,
  title={Transport properties of finite-$\beta$ microturbulence},
  author={Pueschel, MJ and Jenko, F},
  journal={Physics of Plasmas},
  volume={17},
  number={6},
  pages={062307},
  year={2010},
  publisher={American Institute of Physics}
}

@Article{	  romanelli:2014,
  title		= {{JINTRAC: A system of codes for integrated simulation of
		  tokamak scenarios}},
  author	= {Romanelli, M and Corrigan, G and Parail, V and Wiesen, S
		  and Ambrosino, R and Belo, P da Silva Aresta and Garzotti,
		  L and Harting, D and K{\"o}chl, F and Koskela, T and
		  others},
  journal	= {Plasma Fusion Res},
  year		= {2014}
}

@article{ryter:2019,
  title={{Heat transport driven by the ion temperature gradient and electron temperature gradient instabilities in ASDEX Upgrade H-modes}},
  author={Ryter, F and Angioni, C and Dunne, M and Fischer, R and Kurzan, B and Lebschy, A and McDermott, RM and Suttrop, W and Tardini, G and Viezzer, E and others},
  journal={Nuclear Fusion},
  volume={59},
  number={9},
  pages={096052},
  year={2019},
  publisher={IOP Publishing}
}

@article{sertoli:2019,
  title={Measuring the plasma composition in tokamaks with metallic plasma-facing components},
  author={Sertoli, M and Carvalho, Pedro Jorge and Giroud, C and Menmuir, S and Contributors, JET},
  journal={Journal of Plasma Physics},
  volume={85},
  number={5},
  year={2019},
  publisher={Cambridge University Press}
}

@article{smith:2015,
  title={{Electron temperature critical gradient and transport stiffness in DIII-D}},
  author={Smith, Sterling P and Petty, Clinton C and White, Anne E and Holland, Christopher and Bravenec, Ronald and Austin, Max E and Zeng, Lei and Meneghini, Orso},
  journal={Nuclear Fusion},
  volume={55},
  number={8},
  pages={083011},
  year={2015},
  publisher={IOP Publishing}
}

@Article{	  staebler:2007,
  author	= {G.M. Staebler and J.E. Kinsey and R.E. Waltz},
  title		= {A theory-based transport model with comprehensive
		  physics},
  journal	= pop,
  volume	= 14,
  pages		= 055909,
  year		= 2007
}

@Article{	  staebler:2016,
  author	= {G.M. Staebler and J. Candy and N. T. Howard and C. Holland},
  title		= {The role of zonal flows in the saturation of multi-scale
		  gyrokinetic turbulence},
  journal	= pop,
  volume	= {23},
  number	= {},
  pages		= {062518},
  year		= {2016},
  doi = {https://doi.org/10.1063/1.4954905}
}

@Article{	  staebler:2017,
  author	= {G.M. Staebler and N.T. Howard and J. Candy and C. Holland},
  title		= {A model for the coupling of electron and ion scale gyrokinetic turbulence},
  journal	= nf,
  volume	= {57},
  number	= {},
  pages		= {066046},
  year		= {2017},
  doi = {https:://doi.org/10.1088/1741-4326/aa6bee}
}

@article{hein:2010,
  title={Gyrokinetic study of the role of $\beta$ on electron particle transport in tokamaks},
  author={Hein, T and Angioni, C and Fable, E and Candy, J},
  journal={Physics of Plasmas},
  volume={17},
  number={10},
  pages={102309},
  year={2010},
  publisher={American Institute of Physics}
}

@article{stephens:2021,
  title={{Quasilinear gyrokinetic theory: a derivation of QuaLiKiz}},
  author={Stephens, Cole Darin and Garbet, Xavier and Citrin, Jonathan and Bourdelle, Clarisse and van de Plassche, Karel Lucas and Jenko, Frank},
  journal={Journal of Plasma Physics},
  volume={87},
  number={4},
  year={2021},
  publisher={Cambridge University Press}
}

@Misc{stephens:2021b,
 note = {C. Stephens \textit{et al.}, to be submitted to Phys. Plasmas (2021)}
}

@article{watanabe:2005gkv,
  title={Velocity--space structures of distribution function in toroidal ion temperature gradient turbulence},
  author={Watanabe, T-H and Sugama, H},
  journal={Nuclear Fusion},
  volume={46},
  number={1},
  pages={24},
  year={2005},
  publisher={IOP Publishing}
}

@article{white:2019,
  title={Validation of nonlinear gyrokinetic transport models using turbulence measurements},
  author={White, AE},
  journal={Journal of Plasma Physics},
  volume={85},
  number={1},
  year={2019},
  publisher={Cambridge University Press}
}

@Misc{szepesi:2020,
	note    = "Szepesi, G., et al. ''Advanced equilibrium reconstruction for JET with EFIT++'', 47th EPS Conference on Plasma Physics, 2020 ",
}

\end{document}